\documentclass[structabstract,longauth]{aa}  
\usepackage{graphicx}
\usepackage{epstopdf}
\usepackage{subfigure}
\usepackage{longtable}
\usepackage{lscape}
\usepackage{pdflscape}
\usepackage{txfonts}
\usepackage{natbib}
\bibpunct{(}{)}{;}{a}{}{,}
\begin{document}
   \title{Correlations between the stellar, planetary and debris components of exoplanet systems observed by \textit{Herschel}}
   
   \author{Jonathan P. Marshall\inst{1,2}$^{,}$\thanks{email: jonathan.marshall@uam.es}
          \and A.~Moro-Mart\'in\inst{3,4}
          \and C.~Eiroa\inst{1}
          \and G.~Kennedy\inst{5}
          \and A.~Mora\inst{6}
          \and B.~Sibthorpe\inst{7}
          \and J.-F.~Lestrade\inst{8}
          \and J.~Maldonado\inst{1,9}
          \and J.~Sanz-Forcada\inst{10}
          \and M.C.~Wyatt\inst{5}
          \and B.~Matthews\inst{11,12}
          \and J.~Horner\inst{2,13,14}
          \and B.~Montesinos\inst{10}
          \and G.~Bryden\inst{15}
          \and C.~del Burgo\inst{16}
          \and J.S.~Greaves\inst{17}
          \and R.J.~Ivison\inst{18,19}
          \and G.~Meeus\inst{1}
          \and G.~Olofsson\inst{20}
          \and G.L.~Pilbratt\inst{21}
          \and G.J.~White\inst{22,23}
          }

   \institute{Depto. de F\'isica Te\'orica, Universidad Aut\'onoma de Madrid, Cantoblanco, 28049, Madrid, Spain
   \and School of Physics, University of New South Wales, Sydney, NSW 2052, Australia
   \and Department of Astrophysics, Center for Astrobiology, Ctra. de Ajalvir, km 4, Torrejon de Ardoz, 28850, Madrid, Spain
   \and Space Telescope Science Institute, 3700 San Martin Dr, Baltimore, MD 21218, United States
   \and Institute of Astronomy (IoA), University of Cambridge, Madingley Road, Cambridge, CB3 0HA, UK
   \and ESA-ESAC Gaia SOC. P.O. Box 78 E-28691 Villanueva de la Ca{\~n}ada, Madrid, Spain
   \and SRON Netherlands Institute for Space Research, NL-9747 AD Groningen, The Netherlands
   \and Observatoire de Paris, CNRS, 61 Av. de l'Observatoire, F-75014, Paris, France
   \and INAF Observatorio Astronomico di Palermo, Piazza Parlamento 1, 90134 Palermo, Italy
   \and Department of Astrophysics, Centre for Astrobiology (CAB, CSIC-INTA), ESAC Campus, P.O. Box 78, 28691 Villanueva de la Ca\~nada, Madrid, Spain
   \and Herzberg Astronomy \& Astrophysics, National Research Council of Canada, 5071 West Saanich Rd, Victoria, BC V9E 2E7, Canada
   \and University of Victoria, Finnerty Road, Victoria, BC, V8W 3P6, Canada
   \and Australian Centre for Astrobiology, University of New South Wales, Sydney, NSW 2052, Australia
   \and Computational Engineering and Science Research Centre, University of Southern Queensland, Toowoomba, Queensland 4350, Australia
   \and Jet Propulsion Laboratory, California Institute of Technology, Pasadena, CA 91109, USA
   \and Instituto Nacional de Astrof\'\i sica, \'Optica y Electr\'onica, Luis Enrique Erro 1, Sta. Ma. Tonantzintla, Puebla, Mexico
   \and SUPA, School of Physics and Astronomy, University of St. Andrews, North Haugh, St. Andrews KY16 9SS, UK
   \and UK Astronomy Technology Centre, Royal Observatory Edinburgh, Blackford Hill, Edinburgh EH9 3HJ, UK
   \and Institute for Astronomy, University of Edinburgh, Royal Observatory, Blackford Hill, Edinburgh EH9 3HJ, UK
   \and Department of Astronomy, Stockholm University, AlbaNova University Center, Roslagstullsbacken 21, 106 91 Stockholm, Sweden
   \and ESA Astrophysics \& Fundamental Physics Missions Division, ESTEC/SRE-SA,  Keplerlaan 1, NL-2201 AZ Noordwijk, The Netherlands
   \and Department of Physical sciences, The Open University, Walton Hall, Milton Keynes MK7 6AA, UK
   \and Rutherford Appleton Laboratory, Chilton OX11 0QX, UK
   }

   \date{Received ---; accepted ---}

 
  \abstract
   {Stars form surrounded by gas and dust rich protoplanetary discs. Generally, these discs dissipate over a few (3--10) Myr, leaving a faint tenuous debris disc composed of second generation dust produced by the attrition of larger bodies formed in the protoplanetary disc. Giant planets detected in radial velocity and transit surveys of main sequence stars also form within the protoplanetary disc, whilst super-Earths now detectable may form once the gas has dissipated. Our own Solar system, with its eight planets and two debris belts is a prime example of an end state of this process.}
   {The \textit{Herschel} DEBRIS, DUNES and GT programmes observed 37 exoplanet host stars within 25~pc at 70, 100 and 160~$\mu$m with the sensitivity to detect far-infrared excess emission at flux density levels only an order of magnitude greater than that of the Solar system's Edgeworth-Kuiper belt. Here we present an analysis of that sample, using it to more accurately determine the (possible) level of dust emission from these exoplanet host stars and thereafter determine the links between the various components of these exoplanetary systems through statistical analysis.}
   {We have fitted the flux densities measured from recent \textit{Herschel} observations with a simple two parameter ($T_{d}$, $L_{\rm IR}/L_{\star}$) black body model (or to the 3-$\sigma$ upper limits at 100~$\mu$m). From this uniform approach we calculate the fractional luminosity, radial extent, dust temperature and disc mass. We then plotted the calculated dust luminosity or upper limits against the stellar properties, e.g. effective temperature, metallicity, age, and identified correlations between these parameters.}
   {A total of eleven debris discs are identified around the 37 stars in the sample. An incidence of ten cool debris discs around the Sun-like exoplanet host stars (29~$\pm$~9~\%) is consistent with the detection rate found by DUNES (20.2~$\pm$~2.0\%). For the debris disc systems, the dust temperatures range from 20 to 80~K, and fractional luminosities ($L_{\rm IR}/L_{\star}$) between 2.4~$\times$10$^{-6}$ and 4.1~$\times$10$^{-4}$. In the case of non-detections, we calculated typical 3-$\sigma$ upper limits to the dust fractional luminosities of a few~$\times10^{-6}$.}
   {We recover the previously identified correlation between stellar metallicity and hot Jupiter planets in our data set. We find a correlation between the increased presence of dust, lower planet masses and lower stellar metallicities. This confirms the recently identified correlation between cold debris discs and low mass planets in the context of planet formation by core accretion.}

   \titlerunning{Herschel observations of exoplanet host stars}
   \authorrunning{J.P. Marshall, A. Moro-Mart\'in, et al.}
   \maketitle
%
\section{Introduction}

Circumstellar debris discs around main sequence stars are composed of second generation dust produced by the attrition of larger bodies \citep{bp93} which are remnants of primordial protoplanetary discs \citep{hernandez07}. Debris discs can be detected and analyzed based on their excess infrared emission from the constituent dust particles. Around 16.4$^{+2.8}_{-2.9}$~\% of main sequence Sun-like stars have evidence of circumstellar dust emission at 70~$\mu$m with \textit{Spitzer} \citep{trilling08}. From observations of FGK stars by the \textit{Herschel}\footnote{\textit{Herschel} is an ESA space observatory with science instruments provided by European-led Principal Investigator consortia and with important participation from NASA.} DUNES survey an incidence of 20.2~$\pm$~2.0~\% was measured \citep{eiroa13}, whereas the DEBRIS survey measure an incidence of 16.5~$\pm$~2.5~\% (Sibthorpe et al. in prep). 

Many circumstellar discs around Sun-like stars are seen to have two temperature components, which has been interpreted as arising from two distinct belts at different stellocentric radii \citep{chen09,morales11}. The cool discs are more commonly seen and analogous to the Edgeworth-Kuiper belt (EKB) in our own Solar system \citep{greaves10,vitense12}. The EKB's existence has been inferred from the detection of over a thousand Trans-Neptunian Objects\footnote{1258 as of 29th October 2013, see:\\ http://www.minorplanetcenter.net/iau/lists/TNOs.html} , through ground-based surveys and in-situ dust measurement from Voyager 1 and 2 \citep{gurnett97} and New Horizons \citep{poppe10,han11}, although direct observation of the dust emission from the EKB is confounded by the bright foreground thermal emission from the zodiacal dust in the inner Solar system \citep{backman95}. The less commonly seen warm debris disc Asteroid-belt analogues, which are more difficult to observe around other stars due to the larger flux density contribution from the stellar photosphere at mid-infrared wavelengths compared to the dust excess, have been detected around $\sim$~2~\% of Sun-like stars \citep[3/7 FGK stars with 24~$\mu$m excess and $T_{\rm dust}~>~100~$K from a sample of 184,][]{trilling08}.

Exoplanets\footnote{Databases of exoplanet properties are maintained at http://exoplanet.eu and http://exoplanets.org} around Sun-like stars have been identified through radial velocity \citep[e.g. ][]{cnc,aaps,mayor12} or transit surveys e.g. CoRoT \citep{auvergne09}, WASP \citep{wasp}, HAT \citep{hat} and \textit{Kepler} \citep{kepler}. See \citet{perryman11} for a summary of exoplanet detection techniques. The majority of all exoplanet searches have taken place at optical wavelengths, with a sample focus on mature, Sun-like stars as the most suitable candidates for the radial velocity detection technique. A stars are avoided as their atmospheres lack the narrow lines necessary for radial velocity detections through accurate doppler measurements, however these stars are prime candidates for direct imaging surveys, resulting in the detection of several exoplanet systems around debris disc host stars, e.g. Fomalhaut \citep{kalas08}, HR~8799 \citep{marois08}, $\beta$ Pic \citep{lagrange10} and HD~95086 \citep{rameau13}. M stars have likewise been avoided because they exhibit high levels of stellar variability and their emission peaks in the near-infrared, rendering them noisy and faint, although great efforts have been made to overcome these issues due to their sheer number and potential to yield low mass planets through either transit or radial velocity detections \citep[e.g. ][]{reiners10,rodler11,giacobbe12,at12}. This has led to unavoidable bias in the types of stars around which exoplanets are known to exist. Comparative analysis or aggregation of results from radial velocity surveys is further complicated due to both the differences in sensitivity and the variable baseline of observations for the stellar samples, although broad conclusions on e.g. the absence of Jupiter analogues around most nearby stars may be drawn \citep{cumming99,fischer14}. Recent results from an analysis of microlensing surveys suggest that almost all stars may have one or more exoplanets, with low mass planets being much more common than Jupiter mass ones \citep{exoplanets}. On the other hand, long term monitoring from the ground has constrained the likelihood of Jovian planets on long orbits (3--6~AU) to $<$~30\% of the stellar systems surveyed \citep{wittenmyer10}, suggesting that exoplanetary systems very like our own may be rare, a result supported by recent direct imaging searches \citep{janson13}.

Given that planets are the end state of the agglomeration of smaller bodies from dust to planetesimals (neglecting as a detail the capture of an envelope from the protoplanetary disc for gas giants), and debris discs are the result of collisional grinding of these planetesimals into dust, one might expect the presence of planets and debris discs to be correlated. This expectation is strengthened by the direct imaging of several exoplanet systems around debris disc host stars, as previously noted, and indirectly by the structural features observed in many debris discs (warps, off-sets, asymmetries). These features have generally been thought to arise from the gravitational perturbation of exoplanets \citep[see reviews by][]{wyatt08,krivov10,moromartin13}, although remnant gas may offer an alternative explantation in some cases \citep[e.g. ][]{lyra13}. Evidence of disc structures, particularly in the sub-mm, has sometimes weakened upon further scrutiny, e.g. the SCUBA detected 'blob' in Vega's disc \citep{holland98,wyatt03,pietu11,hughes12} or the clumps in the disc of HD~107146 \citep{corder09,hughes11}, requiring that caution be exercised when attributing these structures to unseen planetary bodies. Furthermore, no clear correlation has yet been established between the presence of debris discs and the presence of planets \citep{greaves06,moromartin07b,bryden09,kospal09} and larger scale direct imaging surveys have found little evidence of massive planets around debris disc host stars \citep{wahhaj13,janson13}. On the contrary, \cite{maldonado12} identify a trend between the presence of a debris disc and cool Jupiter exoplanet around a star whilst \cite{wyatt12} identified a possible trend between cool dust and low mass planet host stars.

Direct measurement of the spatial distribution of debris in other stellar systems will reveal whether our own EKB is common or unusual. The EKB and its interaction with the outer planets is thought to have played a significant role in the development of life on Earth, having supplied a significant fraction of the impactors thought to have been involved in the Late Heavy Bombardment \citep{gomes05}, and continues to provide bodies for the short period comet population through interaction with Jupiter \citep{hj09}. It may also have been involved in the hydration of the Earth, a topic which is still widely debated \citep{obrien06,hmpj09,bond10,hj10,izidoro13}. As such, the nature of exo-EKBs may play a vital role in the determination of the habitability of their host systems \citep[e.g. ][]{hj10}, and it is therefore vital that we are able to judge whether the Solar system's architecture and EKB are the norm, or unusual.

The \textit{Herschel} \citep{herschel_ref} Guaranteed Time (GT) debris disc programme and the Open Time Key Programmes DEBRIS\footnote{http://debris.astrosci.ca} \citep[Disc Emission from Bias-free Reconnaissance Infrared Survey; ][]{matthews10} and DUNES\footnote{http://www.mpia-hd.mpg.de/DUNES/} \citep[DUst around NEarby Stars; ][]{eiroa10,eiroa13} have observed nearby Sun-like stars using the PACS \citep[Photodetector Array Camera and Spectrometer,][]{pacs_ref} instrument at far-infrared wavelengths looking for excess emission due to the presence of circumstellar dust discs analogous to the Solar system's EKB \citep{vitense12}. In this work, we have examined all of the stars from the DEBRIS, DUNES and GT programmes currently believed to host exoplanets.

In Section 2, we present the observations used in this work, the data reduction process and the stellar physical parameters to be compared to the dust emission. In Section 3, a summary of the assumptions used to fit models to the new \textit{Herschel} photometry are explained and the calculated disc temperatures and masses are shown. In Section 4, a comparison of the observed disc fractional luminosities (or 3-$\sigma$ upper limits in the case of non-detections) from Section 3 and the stellar properties from Section 2 is presented. Finally, in Section 5, we present our conclusions and recapitulate our findings.
%
\section{Observations and data reduction}

The observations used in this work have been taken from the DEBRIS and DUNES Open Time Key Programmes and the debris disc Guaranteed Time programme. The sample comprises the 14 stars from DEBRIS and 21 stars from DUNES that are main sequence stars within 25~pc of the Sun with radial velocity detected exoplanets. We have also added $\tau$ Ceti and $\epsilon$ Eridani from the Guaranteed Time Key Programme. We have included both $\alpha$ Centauri B and $\epsilon$ Eridani in the sample although there is some doubt over the existence of their respective planets: $\alpha$ Centauri B; \cite{dumusque12,hatzes13}, $\epsilon$ Eridani; \cite{hatzes00,moran04,zechmeister13}. There are no A stars that match the criteria for inclusion; although Fomalhaut and $\beta$ Pictoris both lie within the volume their planets are directly imaged. Likewise around G stars, GJ~504 hosts a cold Jovian planet \citep{kuzuhara13} and HN Peg has a borderline brown dwarf/Jovian companion \citep{faherty10}. The range of spectral types represented in the sample is therefore F to M. Several exoplanet host stars have not been included, e.g. HD~136352, HD~147513 and HD~190360 \citep{wyatt12}, despite lying within the volume explored by DEBRIS and DUNES, as they all lie towards regions of bright far-infrared background contamination from Galactic emission and were therefore omitted from observation by the \textit{Herschel} programmes.

\textit{Herschel} PACS 70/160 and/or 100/160 scan map observations were taken of all stars except 51 Peg, which was observed in chop-nod mode with the 100/160 channel combination during the Science Demonstration Phase. The observation parameters for the DEBRIS and DUNES targets were slightly different: for DEBRIS targets, each scan map consisted of two repetitions of eight scan legs of 3\arcmin~length, with a 4\arcsec~separation between legs, taken at the medium slew speed (20\arcsec~per second) whereas DUNES targets had the same parameters except they used ten legs per scan map and the number of repetitions was dictated by the requirement to detect the stellar photosphere at 100~$\mu$m with a signal to noise ratio (SNR) $\geq~5$. The PACS 100/160 scan map observations of q$^{1}$ Eri were observed as part of the calibration effort during the SDP phase using scans with 16 legs of 3.9\arcmin~with 4\arcsec~separation between the legs, also at the medium slew speed. For the GT programme, the targets were observed twice with each scan map consisting of 11 scan legs of 7.4\arcmin~ length, with a 38\arcsec~ separation between legs, at the medium slew speed. Each of the GT scans was repeated 11 times. In all cases, each target was observed at two array orientation angles (70\degr~ and 110\degr~ for DEBRIS and DUNES, 45\degr~ and 135\degr~ for GT) which were combined into a final mosaic to improve noise suppression and assist in the removal of instrumental artefacts and glitches. Several of the targets presented in this work have been observed by \textit{Herschel} SPIRE \citep[Spectral and Photometric Imaging REceiver; ][]{spire_ref,spire_cal}, but due to the sparse coverage of the sample with that instrument (4/37 targets) we decided to focus here on the PACS photometry to ensure consistency in the data set used for the sample analysis. A summary of all observations is presented in Table \ref{obs_log}.

\begin{table}[ht]
\centering
\caption{Summary of \textit{Herschel} PACS observations of the target stars.}
\begin{tabular}{lrrcr}
\hline\hline
\multicolumn{1}{c}{Target} & \multicolumn{1}{c}{PACS} & \multicolumn{1}{c}{Obs. IDs} & \multicolumn{1}{c}{OD$^{a}$} & \multicolumn{1}{c}{OT$^{b}$} \\
       & \multicolumn{1}{c}{[$\lambda$ $\mu$m]} & \multicolumn{1}{c}{[1342\ldots]} &          & \multicolumn{1}{c}{[s]} \\
\hline
HD~1237 & 100/160 & 204272/73 & 484 & 576 \\
HD~3651 & 70/160 & 213242/43 & 621 & 360\\
         & 100/160 & 213244/45 & 621 & 1080\\
HD~4308 & 70/160 & 212704/05 & 612 & 180\\
         & 100/160 & 212706/07 & 612 & 1440\\
$\upsilon$~And & 100/160 & 223326/27 & 777 & 360\\
q$^{1}$~Eri & 70/160 & 212838/39 & 614 & 360\\
         & 100/160 & 187139/40 & 191 & 768 \\
$\tau$~Ceti & 70/160 & 213575/76 & 628 & 5566 \\
GJ 86 & 70/160 & 214175/76 & 639 & 180\\
          & 100/160 & 214177/78 & 639 & 720\\
$\iota$~Hor & 100/160 & 214196/97 & 640 & 576\\
HD~19994 & 100/160 & 216129/30 & 661 & 1440\\
HD~20794 & 100/160 & 216456/57 & 675 & 1220 \\
$\epsilon$~Eri & 70/160 & 216123/24 & 661 & 5566 \\
HD~33564 & 100/160 & 219019/20 & 705 & 720\\
HD~39091 & 100/160 & 205212/13 & 502 & 576 \\
HD~40307 & 70/160 & 203666/67 & 471 & 180\\
         & 100/160 & 203668/69 & 471 & 1440\\
HD~69830 & 100/160 & 196125/26 & 361 & 900\\
55~Cnc & 100/160 & 208504/05 & 545 & 720\\
47~UMa & 100/160 & 198845/46 & 403 & 360\\
HD~99492 & 100/160 & 212056/57 & 582 & 576 \\
HD~102365 & 100/160 & 202240/41 & 450 & 576 \\
61~Vir & 100/160 & 202551/52 & 454 & 576 \\
70~Vir & 100/160 & 213093/94 & 615 & 1440\\
$\tau$~Boo & 100/160 & 213081/82 & 615 & 360\\
$\alpha$~Cen B & 100/160 & 224848/49 & 807 & 360 \\
GJ~581 & 100/160 & 202568/69 & 454 & 576 \\
$\rho$~CrB & 100/160 & 215376/77 & 662 & 1080\\
14~Her & 100/160 & 205996/97 & 511 & 1440\\
HD~154345 & 70/160 & 193509/10 & 324 &  1080\\
          & 100/160 & 193511/12 & 324 & 1440\\
$\mu$~Ara & 100/160 & 215572/73 & 663 & 360\\
HD~176051 & 100/160 & 205038/39 & 497 & 576 \\
16~Cyg B & 100/160 & 198501/02 & 400 & 576 \\
HD~189567 & 100/160 & 208851/52 & 547 & 576 \\
HD~192310 & 100/160 & 208466/67 & 545 & 576 \\
GJ~832 & 100/160 & 208845/46 & 547 & 576 \\
HD~210277 & 100/160 & 211126/27 & 579 & 1440\\
GJ~876 & 100/160 & 198521/22 & 400 & 576 \\
51~Peg & 100/160 & 187255 & 198 & 682\\
HD~217107 & 100/160 & 198515/16 & 400 & 576 \\
\hline
\end{tabular}
\tablefoot{(a) Operational Day; (b) On-source integration time.}
\label{obs_log}
\vspace{-0.20in}
\end{table}

All observations were reduced interactively using version 8.1.0 of the Herschel Interactive Processing Environment \citep[HIPE, ][]{ott10} using PACS calibration version 32 and the standard scripts supplied with HIPE. Individual PACS scans were processed with a high pass filter to remove background structure, using high pass filter widths of 15 frames at 70~$\mu$m, 20 frames at 100~$\mu$m and 25 frames at 160~$\mu$m, equivalent to spatial scales of 62\arcsec, 82\arcsec~and 102\arcsec. In the case of $\epsilon$ Eridani, with its much larger disc ($\sim~$1\arcmin~in diameter), a high pass filter width of 50 frames, equivalent to 202\arcsec, at both 70 and 160~$\mu$m was adopted to avoid removal of disc flux by the filtering. For the filtering process, regions of the map where the pixel brightness exceeded a threshold defined as twice the standard deviation of the non-zero elements in the map were masked from the high pass filter task. The two individual scans of each target were mosaicked to reduce sky noise and suppress the striping due to detector scanning. Final image scales were 1\arcsec~per pixel at 70~$\mu$m and 100~$\mu$m and 2\arcsec~per pixel at 160~$\mu$m compared to native instrument pixel sizes of 3.2\arcsec~at 70~$\mu$m and 100~$\mu$m and 6.4\arcsec~at 160~$\mu$m.

Point source flux densities were measured with aperture radii of 4\arcsec, 5\arcsec~and 8\arcsec~at 70~$\mu$m, 100~$\mu$m and 160~$\mu$m to maximize the snr of the source \citep{eiroa13}. The mean FWHM of the PACS instrument is 5.61\arcsec, 6.79\arcsec~and 11.36\arcsec~in the three bands. Extended sources, identified by comparison of a 2D Gaussian fit to the source profile to the PSF FWHM in each band, were measured with varying aperture radii depending on the disc extent. The sky background level and rms noise for each target were estimated from the mean and standard deviation of the total flux density in 25 boxes of dimensions 7$\times$7 pixels at 70~$\mu$m, 9$\times$9 pixels at 100~$\mu$m and 7$\times$7 pixels at 160~$\mu$m, chosen to match the aperture size of point sources. The boxes were placed randomly around the central area of the mosaic within a region 30\arcsec~to 60\arcsec~from the source position, or 90\arcsec--120\arcsec~in the case of $\epsilon$ Eri, to avoid both the central source, and edges of the maps where the noise increases.

\begin{landscape}
\begin{table}[ht!!]
\centering
\caption{\textit{Spitzer} MIPS70 and \textit{Herschel} PACS photometry along with photospheric estimates for the exoplanet host stars.} \label{photometry}
\begin{tabular}{lcccccccccccc}
\hline\hline
Name & $F_{\rm MIPS}$[70] & $F_{\rm PACS}$[70] & $F_{\star}$[70] & $F_{\rm PACS}$[100] & $F_{\star}$[100] & $F_{\rm PACS}$[160] & $F_{\star}$[160] & $\chi_{100}$ & $\chi_{160}$ & $L_{\rm IR}/L_{\star} $ & $R_{\rm disc}$ & $T_{\rm disc}$ \\
     & [mJy] & [mJy] & [mJy] & [mJy] & [mJy] & [mJy] & [mJy] &  &  & [$\times10^{-6}$] & [AU] & [K] \\
\hline
HD~1237 & \phantom{0}11.2~$\pm$~\phantom{0}2.5 & \ldots & \phantom{0}9.56 & \phantom{00}4.57~$\pm$~\phantom{0}1.19 & \phantom{0}4.69 & \phantom{0}-0.72~$\pm$~\phantom{0}3.36 & \phantom{0}1.81 & -0.1 & -0.8 & $<$6.4 & (45.3) & \\
HD~3651 & \phantom{0}14.9~$\pm$~\phantom{0}5.7 & \phantom{0}21.93~$\pm$~\phantom{0}1.59 & 22.90 & \phantom{00}8.21~$\pm$~\phantom{0}1.21 & 11.22 & \phantom{00}3.76~$\pm$~\phantom{0}3.43 & \phantom{0}4.38 & -2.5 & -0.2 & $<$4.6 & (41.1) & \\
HD~4308 & \phantom{00}5.2~$\pm$~\phantom{0}4.4 & \phantom{00}7.97~$\pm$~\phantom{0}0.93 & \phantom{0}8.72 & \phantom{00}5.23~$\pm$~\phantom{0}1.00 & \phantom{0}4.27 & \phantom{00}1.63~$\pm$~\phantom{0}1.63 & \phantom{0}1.67 & 1.0 & -0.0 & $<$5.8 & (56.8) & \\
$\upsilon$ And & \phantom{0}55.7~$\pm$~\phantom{0}6.3 & \ldots & 60.22 & \phantom{0}32.78~$\pm$~\phantom{0}1.67 & 29.51 & \phantom{0}16.49~$\pm$~\phantom{0}2.64 & 11.53 & 2.0 & 1.9 & $<$3.1 & (103.5) &  \\
\textbf{q$^{1}$ Eri} & 863.4~$\pm$~58.7 & 896.20~$\pm$~26.90 & 17.48 & 897.10~$\pm$~26.90 & \phantom{0}8.56 & 635.90~$\pm$~31.80 & \phantom{0}3.35 & 33.0 & 19.9 & 405.0 & 25.7 & 60.9\\
\textbf{$\tau$ Ceti}$^{a}$ & \ldots & 301.0~$\pm$~15.0 & 183.50 & \ldots & 89.92 & 103.0~$\pm$~5.0 & 32.20 & 4.8 & 14.2 & 7.8 & 12.2 & 67.5 \\
GJ~86 & \phantom{00}6.9~$\pm$~\phantom{0}6.8 & \phantom{0}14.28~$\pm$~\phantom{0}1.66 & 18.86 & \phantom{00}7.07~$\pm$~\phantom{0}1.30 & \phantom{0}9.24 & \phantom{00}3.78~$\pm$~\phantom{0}2.49 & \phantom{0}3.61 & -1.7 & 0.1 & $<$5.3 & (36.0) &    \\
$\iota$ Hor & \phantom{0}20.1~$\pm$~\phantom{0}4.1 & \ldots & 18.82 & \phantom{00}7.34~$\pm$~\phantom{0}1.25 & \phantom{0}9.22 & \phantom{00}4.29~$\pm$~\phantom{0}3.10 & \phantom{0}3.55 & -1.5 & 0.2 & $<$3.1 & (69.6) &   \\
\textbf{HD~19994} & \phantom{0}42.5~$\pm$~\phantom{0}3.5 & \ldots & 26.43 & \phantom{0}39.45~$\pm$~\phantom{0}1.75 & 12.95 & \phantom{0}31.75~$\pm$~\phantom{0}1.94 & \phantom{0}5.06 & 15.1 & 13.8 & 5.4 & 91.1 & 40.8\\ 
\textbf{HD~20794}$^{b}$ & \phantom{0}94.0~$\pm$~13.6 & \phantom{0}97.0~$\pm$~\phantom{0}5.5 & 83.62 & \phantom{0}56.01~$\pm$~\phantom{0}2.86 & 40.97 & \phantom{0}23.37~$\pm$~\phantom{0}2.42 & 15.79 & 5.3 & 3.1 & 2.4 & 10.8 & 76.5\\
\textbf{$\epsilon$ Eri}$^{a}$ & 1688.0~$\pm$~10.0 & 1599.5~$\pm$~80.0 & 187.60 & \ldots & 91.90 & 1120.8~$\pm$~56.0 & 35.90 & 9.5 & 19.4 & 107.6 & 13.4 & 57.9 \\
HD~33564 & \phantom{0}32.6~$\pm$~\phantom{0}5.5 & \ldots & 22.04 & \phantom{0}12.02~$\pm$~\phantom{0}2.24 & 10.80 & \phantom{00}7.30~$\pm$~\phantom{0}2.70 & \phantom{0}4.16 & 0.5 & 1.2 & $<$4.0 & (101.4) &  \\
HD~39091 & \phantom{0}23.4~$\pm$~\phantom{0}2.9 & \ldots & 16.15 & \phantom{0}12.26~$\pm$~\phantom{0}1.84 & \phantom{0}7.91 & \phantom{00}5.59~$\pm$~\phantom{0}2.87 & \phantom{0}3.05 & 2.4 & 0.9 & $<$4.8 & (69.9) &   \\
\textbf{HD~40307}$^{b}$ & \phantom{0}17.2~$\pm$~\phantom{0}4.9 & \phantom{0}14.60~$\pm$~\phantom{0}1.43 & 10.01 & \phantom{00}8.05~$\pm$~\phantom{0}0.95 & \phantom{0}4.90 & \phantom{00}8.02~$\pm$~\phantom{0}1.50 & \phantom{0}1.92 & 3.3 & 4.1 & 4.3 & 24.0 & 39.9\\
\textbf{HD~69830}$^{c}$ & \phantom{0}15.1~$\pm$~\phantom{0}2.4 & \ldots & 19.84 & \phantom{00}9.33~$\pm$~\phantom{0}0.72 & \phantom{0}9.72 & \phantom{00}2.01~$\pm$~\phantom{0}2.01 & \phantom{0}3.80 & -0.5 & -0.9 & $<$4.6 & (43.5) & \\
55 Cnc & \phantom{0}19.8~$\pm$~\phantom{0}4.4 & \ldots & 19.66 & \phantom{00}8.59~$\pm$~\phantom{0}0.87 & \phantom{0}9.64 & \phantom{00}4.29~$\pm$~\phantom{0}1.92 & \phantom{0}3.76 & -1.2 & 0.3 & $<$4.8 & (43.8) &  \\
47 UMa & \phantom{0}31.4~$\pm$~\phantom{0}4.2 & \ldots & 30.26 & \phantom{0}13.52~$\pm$~\phantom{0}1.16 & 14.83 & \phantom{00}2.82~$\pm$~\phantom{0}2.82 & \phantom{0}5.79 & -1.1 & -1.1 & $<$3.4 & (71.4) &  \\
HD~99492 & \phantom{00}7.5~$\pm$~\phantom{0}6.3 & \ldots & \phantom{0}6.79 & \phantom{00}5.68~$\pm$~\phantom{0}1.03 & \phantom{0}3.33 & \phantom{00}4.57~$\pm$~\phantom{0}3.96 & \phantom{0}1.30 & 2.3 & 0.8 & $<$10.6 & (32.3) &   \\
HD~102365 & \phantom{0}34.0~$\pm$~\phantom{0}8.5 & \ldots & 41.43 & \phantom{0}20.89~$\pm$~\phantom{0}1.43 & 20.30 & \phantom{00}7.53~$\pm$~\phantom{0}2.99 & \phantom{0}7.83 & 0.4 & -0.1 & $<$4.2 & (51.7) &   \\
\textbf{61 Vir} & 185.6~$\pm$~16.6 & 198.00~$\pm$~10.34 & 49.77 & 156.64~$\pm$~\phantom{0}8.63 & 24.39 & 131.30~$\pm$~\phantom{0}7.13 & \phantom{0}9.40 & 15.3 & 17.1 & 27.6 & 22.4 & 56.1 \\
\textbf{70 Vir} & \phantom{0}79.0~$\pm$~\phantom{0}8.1 & \ldots & 42.07 & \phantom{0}40.73~$\pm$~\phantom{0}2.14 & 20.61 & \phantom{0}26.97~$\pm$~\phantom{0}1.39 & \phantom{0}8.05 & 9.4 & 13.6 & 4.8 & 49.3 & 52.1\\
$\tau$ Boo & \phantom{0}33.2~$\pm$~\phantom{0}6.8 & \ldots & 38.07 & \phantom{0}16.97~$\pm$~\phantom{0}1.37 & 18.65 & \phantom{00}6.24~$\pm$~\phantom{0}2.50 & \phantom{0}7.29 & -1.2 & -0.4 & $<$2.7 & (98.8) &  \\
$\alpha$ Cen B & 1012.0~$\pm$~667.8 & 1490.0~$\pm$~280.0 & 1510.00 & 670.0~$\pm$~37.0 & 740.00 & 210.0~$\pm$~60.0 & 289.00 & -1.65 & -1.26 & $<$0.9 & (37.6) & \\
\textbf{GJ~581} & \phantom{0}15.9~$\pm$~\phantom{0}4.7 & \phantom{0}18.90~$\pm$~\phantom{0}1.40 & \phantom{0}7.06 & \phantom{0}10.96~$\pm$~\phantom{0}1.08 & \phantom{0}3.46 & \phantom{0}18.04~$\pm$~\phantom{0}3.50 & \phantom{0}1.33 & 12.0 & 4.2 & 91.0 & 5.0 & 41.0 \\
$\rho$ CrB & \phantom{0}29.6~$\pm$~\phantom{0}5.4 & \ldots & 23.06 & \phantom{0}10.59~$\pm$~\phantom{0}1.14 & 11.30 & \phantom{00}1.83~$\pm$~\phantom{0}1.83 & \phantom{0}4.41 & -0.6 & -1.4 & $<$3.8 & (74.5) &  \\
14 Her & \phantom{0}10.6~$\pm$~\phantom{0}2.5 & \ldots & 10.76 & \phantom{00}3.91~$\pm$~\phantom{0}0.77 & \phantom{0}5.27 & \phantom{00}1.61~$\pm$~\phantom{0}1.61 & \phantom{0}2.06 & -1.8 & -0.3 & $<$4.8 & (45.6) &  \\
HD~154345 &\phantom{0}11.0~$\pm$~\phantom{0}9.0 & \phantom{00}6.78~$\pm$~\phantom{0}1.32 & \phantom{0}8.16 & \phantom{00}3.94~$\pm$~\phantom{0}0.66 & \phantom{0}4.00 & \phantom{00}1.93~$\pm$~\phantom{0}1.93 & \phantom{0}1.56 & -0.1 & 0.2 & $<$5.5 & (44.3) &  \\
$\mu$ Ara & \phantom{0}31.0~$\pm$~\phantom{0}7.8 & \ldots & 29.08 & \phantom{0}15.46~$\pm$~\phantom{0}2.39 & 14.25 & \phantom{00}7.06~$\pm$~\phantom{0}1.72 & \phantom{0}5.57 & 0.5 & 0.9 & $<$4.8 & (76.2) &  \\
HD~176051 & \phantom{0}31.9~$\pm$~\phantom{0}5.5 & \ldots & 27.78 & \phantom{0}12.30~$\pm$~\phantom{0}1.32 & 13.61 & \phantom{0}-0.84~$\pm$~\phantom{0}3.45 & \phantom{0}5.25 & -1.0 & -1.8 & $<$3.7 & (71.5) & \\
16 Cyg B & \phantom{0}30.7~$\pm$~\phantom{0}7.8 & \ldots & 11.25 & \phantom{0}11.10~$\pm$~\phantom{0}4.40 & \phantom{0}5.51 & \phantom{00}0.00~$\pm$~\phantom{0}3.06 & \phantom{0}2.12 & 1.3 & -0.7 & $<$10.9 & (63.6) &  \\
HD~189567 & \phantom{0}19.0~$\pm$~\phantom{0}3.4 & \phantom{0}10.44~$\pm$~\phantom{0}2.20 & 12.88 & \phantom{00}5.71~$\pm$~\phantom{0}1.00 & \phantom{0}6.31 & \phantom{00}1.85~$\pm$~\phantom{0}1.85 & \phantom{0}2.46 & -0.6 & -0.3 & $<$4.5 & (57.2) &   \\
HD~192310 & \phantom{0}19.1~$\pm$~\phantom{0}5.6 & \ldots & 29.81 & \phantom{0}15.52~$\pm$~\phantom{0}1.26 & 14.61 & \phantom{00}7.34~$\pm$~\phantom{0}3.30 & \phantom{0}5.63 & 0.7 & 0.5 & $<$5.8 & (35.8) &  \\
GJ~832 & \phantom{0}19.5~$\pm$~\phantom{0}4.1 & \ldots & 19.66 & \phantom{0}10.93~$\pm$~\phantom{0}1.82 & \phantom{0}9.63 & \phantom{00}0.00~$\pm$~\phantom{0}3.59 & \phantom{0}3.72 & 0.7 & -1.0 & $<$19.0 & (9.9) &  \\
\textbf{HD~210277}$^{b}$ & \phantom{00}7.9~$\pm$~\phantom{0}2.2 & \ldots & \phantom{0}9.51 & \phantom{00}8.50~$\pm$~\phantom{0}1.00 & \phantom{0}4.66 & \phantom{0}12.40~$\pm$~\phantom{0}1.60 & \phantom{0}1.82 & 3.8 & 6.6 & 5.1 & 154.9 & 22.3\\
GJ~876 & \phantom{00}9.5~$\pm$~\phantom{0}3.1 & \ldots & 14.58 & \phantom{00}7.72~$\pm$~\phantom{0}1.48 & \phantom{0}7.14 & \phantom{00}4.30~$\pm$~\phantom{0}3.14 & \phantom{0}2.75 & 0.4 & 0.5 & $<$23.7 & (6.4) &  \\
51 Peg & \phantom{0}28.1~$\pm$~\phantom{0}4.9 & \ldots & 21.88 & \phantom{0}11.49~$\pm$~\phantom{0}2.65 & 10.72 & \phantom{00}5.00~$\pm$~\phantom{0}5.00 & \phantom{0}4.19 & 0.3 & 0.2 & $<$5.5 & (66.0) &  \\
HD~217107 & \phantom{0}10.9~$\pm$~\phantom{0}5.3 & \ldots & 12.37 & \phantom{00}4.69~$\pm$~\phantom{0}0.90 & \phantom{0}6.10 & \phantom{00}1.69~$\pm$~\phantom{0}3.20  & \phantom{0}2.30 & -1.6 & -0.2 & $<$4.1 & (60.9) &  \\
\hline
\end{tabular}
\tablefoot{\textit{Spitzer} photometry from \cite{eiroa13} or provided by Geoff Bryden (priv. comm.). Debris disc stars are highlighted in bold; (a) $\chi_{100}$ estimated from model fit to 70 and 160~$\mu$m fluxes; (b) \textit{Herschel} discovered debris disc; (c) Warm debris disc.}
\end{table}
\end{landscape}

\begin{landscape}
\begin{table}[ht!!]
\centering
\caption{Stellar and planetary parameters of the exoplanet host stars sample.}
\label{star_params}
\begin{tabular}{llrccccccclll}
\hline\hline
\multicolumn{1}{c}{Name} & \multicolumn{1}{c}{Spectral} & \multicolumn{1}{c}{Distance} &  $L_{\star}$ & $T_{\star}$ & [Fe/H] & \multicolumn{1}{c}{Ca~{\sc ii} Age}  & \multicolumn{1}{c}{X-ray Age} & \multicolumn{1}{c}{$N_{\rm Planet}$$^{a}$}  & \multicolumn{1}{c}{$M~\sin i$$^{b}$} & \multicolumn{1}{c}{$a$$^{c}$} & \multicolumn{1}{c}{$e_{\rm Planet}$$^{d}$} & Ref. \\
       &   \multicolumn{1}{c}{Type}   &   \multicolumn{1}{c}{[pc]}   & \multicolumn{1}{c}{[$L_{\odot}$]} & \multicolumn{1}{c}{[K]} &        & \multicolumn{1}{c}{[Gyr]} & \multicolumn{1}{c}{[Gyr]} &   & \multicolumn{1}{c}{[$M_{\rm Jup}$]} & \multicolumn{1}{c}{[AU]} & & \\
\hline
HD~1237 & G8~V & 17.49~$\pm$~0.09 & 0.643 & 5514 & 0.07 & 0.3 & 0.4 & 1 & 3.37~$\pm$~0.09 & 0.4947~$\pm$~0.0083 & 0.511~$\pm$~0.017 & \phantom{0}1 \\ %
HD~3651 & K0~V & 11.06~$\pm$~0.04 & 0.529 & 5204 & 0.16 & 6.4 & 4.5 & 1 & 0.229~$\pm$~0.015 & 0.2947~$\pm$~0.0049 & 0.63~$\pm$~0.04 & \phantom{0}2 \\ %
HD~4308 & G3~V & 22.06~$\pm$~0.16 & 1.012 & 5670 & -0.24 & 4.0 & 4.5 & 1 & 0.0405~$\pm$~0.005 & 0.118~$\pm$~0.0009 & 0.27~$\pm$~0.12 & \phantom{0}3 \\ %
$\upsilon$ And & F8~V & 13.49~$\pm$~0.03 & 3.363 & 6155 & 0.10 & 7.3 & 2.9 & 3 & 4.12~$\pm$~0.165 & 2.525~$\pm$~0.042 & 0.013~$\pm$~0.016 & \phantom{0}4 \\ %
q$^{1}$ Eri & F8~V & 17.43~$\pm$~0.08 & 1.523 & 6155 & -0.04 & 1.2 & 0.8 & 1 & 0.93~$\pm$~0.18 & 2.03~$\pm$~0.15 & 0.1~$\pm$~0.01 & \phantom{0}5 \\ %
$\tau$ Ceti & G8~V & 3.65~$\pm$~0.01 & 0.526 & 5312 & -0.43 & 8.5 & \ldots & 5 & 0.02~$\pm$~0.01 & 1.35~$\pm$~0.09 & 0.16~$^{+0.22}_{-0.16}$ & \phantom{0}6 \\
GJ~86 & K0~V & 10.78~$\pm$~0.04 & 0.406 & 5165 & -0.22 & 1.5 & 3.6 & 1 & 4.00~$\pm$~0.137 & 0.1142~$\pm$~0.0019 & 0.0416~$\pm$~0.0072 & \phantom{0}7 \\ %
$\iota$ Hor & F8~V & 17.17~$\pm$~0.06 & 1.520 & 6227 & 0.19 & 1.6 & 0.5 & 1 & 2.05~$\pm$~0.20 & 0.924~$\pm$~0.016 & 0.14~$\pm$~0.13 & \phantom{0}8 \\ %
HD~19994 & F8~V & 22.58~$\pm$~0.14 & 3.848 & 6187 & 0.21 & 5.67 & 1.2 & 1 & 1.33~$\pm$~0.11 & 1.306~$\pm$~0.022 & 0.266~$\pm$~0.014 & \phantom{0}9 \\ %
HD~20794 & G8~V & 6.04~$\pm$~0.01 & 0.663 & 5413 & -0.34 & 6.22 & 10.4 & 3 & 0.0148~$\pm$~0.0018 & 0.3498~$\pm$~0.0058 & 0 & 10 \\ %
$\epsilon$ Eri & K2~V & 3.22~$\pm$~0.01 & 0.425 & 5061 & -0.08 & 0.8 & \ldots & 1 & 1.05~$\pm$~0.19 & 3.38~$\pm$~0.32 & 0.25~$\pm$~0.23 & 11 \\
HD~33564 & F6~V & 20.88~$\pm$~0.09 & 3.224 & 6307 & 0.10 & 5.7 & 1.2 & 1 & 9.13~$\pm$~0.37 & 1.124~$\pm$~0.020 & 0.34~$\pm$~0.02 & 12 \\ %
HD~39091 & G1~IV & 18.32~$\pm$~0.07 & 1.535 & 6003 & 0.09 & 5.1 & 3.2 & 1 & 10.09~$\pm$~0.38 & 3.35~$\pm$~0.10 & 0.6405~$\pm$~0.0072 & 13 \\ %
HD~40307 & K3~V & 13.00~$\pm$~0.06 & 0.245 & 4979 & -0.31 & 7.3 & 6.7 & 3 & 0.0281~$\pm$~0.0002 & 0.1324~$\pm$~0.0022 & 0 & 14 \\ %
HD~69830 & K0~V & 12.49~$\pm$~0.05 & 0.595 & 5405 & -0.04 & 6.4 & 3.3 & 3 & 0.0563~$\pm$~0.0052 & 0.627~$\pm$~0.012 & 0.10~$\pm$~0.04 & 15 \\ %
55 Cnc & G8~V & 12.34~$\pm$~0.11 & 0.602 & 5295 & 0.36 & 8.4 & 11.2 & 5 & 3.835~$\pm$~0.08 & 5.475~$\pm$~0.094 & 0.004~$\pm$~0.003 & \phantom{0}4 \\ %
47 UMa & G0~V & 14.06~$\pm$~0.05 & 1.600 & 5908 & 0.03 & 4.9 & \ldots & 3 & 2.546~$\pm$~0.096 & 3.570~$\pm$~0.111 & 0.032~$\pm$~0.014 & 16 \\ %
HD~99492 & K2~V & 17.96~$\pm$~0.46 & 0.327 & 4810 & 0.26 & 4.0 & 13.0 & 2 & 0.36~$\pm$~0.02 & 5.4~$\pm$~0.1 & 0.254~$\pm$~0.092 & 17 \\ %
HD~102365 & G3/G5~V & 9.22~$\pm$~0.02 & 0.839 & 5630 & -0.28 & 5.7 & 8.83 & 1 & 0.051~$\pm$~0.008 & 0.4633~$\pm$~0.0078 & 0.34~$\pm$~0.14 & 18 \\ 
61 Vir & G5~V & 8.56~$\pm$~0.02 & 0.835 & 5646 & -0.02 & 6.6 & 8.0 & 3 & 0.0716~$\pm$~0.0093 & 0.475~$\pm$~0.008 & 0.12~$\pm$~0.11 & 19 \\ %
70 Vir & G5~V & 17.98~$\pm$~0.08 & 2.989 & 5513 & -0.07 & 7.9 & 5.6 & 1 & 7.46~$\pm$~0.25 & 0.4836~$\pm$~0.0081 & 0.4007~$\pm$~0.0035 & 20 \\ %
$\tau$ Boo & F7~V & 15.62~$\pm$~0.05 & 3.062 & 6376 & 0.26 & 4.8 & 0.4 & 1 & 5.95~$\pm$~0.28 & 0.046 & 0.023~$\pm$~0.015 & \phantom{0}4 \\ %
$\alpha$ Cen B & K1~V & 1.25~$\pm$~0.04 & 0.444 & 5178 & 0.15 & 5.2 & 4.27 & 1 & 0.00356~$\pm$~0.00028 & 0.0419~$\pm$~0.0007 & 0 & 21 \\
GJ~581 & M5~V & 6.27~$\pm$~0.09 & 0.012 & 3315 & -0.02 & 2.00 & 7.0 & 4$^{e}$ & 0.0499~$\pm$~0.0023 & 0.2177~$\pm$~0.0047 & 0.32~$\pm$~0.09 & 22 \\ %
$\rho$ CrB & G2~V & 17.24~$\pm$~0.08 & 1.742 & 5834 & -0.19 & 4.3 & $<$1.09 & 1 & 1.064~$\pm$~0.053 & 0.2257~$\pm$~0.0038 & 0.057~$\pm$~0.028 & 23 \\ %
14 Her & K0~V & 17.57~$\pm$~0.10 & 0.653 & 5336 & 0.43 & 6.9 & 7.5 & 1 & 5.21~$\pm$~0.30 & 2.934~$\pm$~0.084 & 0.369~$\pm$~0.005 & 24 \\ %
HD~154345 & G8~V & 18.58~$\pm$~0.11 & 0.617 & 5488 & -0.07 & 3.8 & 5.6 & 1 & 0.957~$\pm$~0.061 & 4.21~$\pm$~0.11 & 0.044~$^{+0.046}_{-0.044}$ & 25 \\ %
$\mu$ Ara & G5~V & 15.51~$\pm$~0.07 & 1.821 & 5787 & 0.29 & 7.5 & \ldots & 4 & 1.89~$\pm$~0.22 & 5.34~$\pm$~0.40 & 0.172~$\pm$~0.040 & 26 \\ %
HD~176051 & G0~V & 14.87~$\pm$~0.08 & 1.603 & 5840 & -0.11 & 8.1 & 1.1 & 1 & 1.50~$\pm$~0.30 & 1.76~$\pm$~0.07 & 0 & 27 \\ %
16 Cyg B & G5~V & 21.21~$\pm$~0.12 & 1.271 & 5772 & 0.08 & 7.4 & $<$10.3 & 1 & 1.640~$\pm$~0.083 & 1.660~$\pm$~0.028 & 0.681~$\pm$~0.017 & 28 \\ %
HD~189567 & G2~V & 17.73~$\pm$~0.14 & 1.027 & 5735 & -0.22 & 4.1 & \ldots & 1 & 0.0316~$\pm$~0.0034 & 0.1099~$\pm$~0.0018 & 0.23~$\pm$~0.14 & 29 \\ %
HD~192310 & K3~V & 8.91~$\pm$~0.02 & 0.407 & 5105 & -0.03 & 7.5 & 2.9 & 2 & 0.0736~$\pm$~0.0098 & 1.184~$\pm$~0.024 & 0.13~$\pm$~0.04 & \textbf{30} \\ %
GJ~832 & M1~V & 7.86~$\pm$~0.12 & 0.031 & 3695 & -0.31 & \ldots & 9.2 & 1 & 0.644~$\pm$~0.075 & 3.40~$\pm$~0.15 & 0.12~$\pm$~0.11 & 31 \\ %
HD~210277 & G0~V & 21.56~$\pm$~0.22 & 1.002 & 5540 & 0.22 & 6.8 & \ldots & 1 & 1.273~$\pm$~0.005 & 1.131~$\pm$~0.019 & 0.476~$\pm$~0.017 & 32 \\ %
GJ~876 & M5~V & 4.69~$\pm$~0.05 & 0.013 & 3473 & 0.19 & 6.5 & \ldots & 4 & 2.2756~$\pm$~0.0045 & 0.3343~$\pm$~0.0013 & 0.207~$\pm$~0.055 & 33 \\ %
51 Peg & G5~V & 15.61~$\pm$~0.09 & 1.368 & 5791 & 0.20 & 8.0 & 13.6 & 1 & 0.461~$\pm$~0.016 & 0.05211~$\pm$~0.00087 & 0.013~$\pm$~0.012 & 34 \\ %
HD~217107 & G8~IV & 19.86~$\pm$~0.15 & 1.162 & 5646 & 0.4 & 8.1 & \ldots & 2 & 2.62~$\pm$~0.15 & 5.33~$\pm$~0.02 & 0.1276~$\pm$~0.0052 & 35 \\ %
\hline
\end{tabular}
\tablefoot{$^{a}$Number of known exoplanets. $^{b}$Projected (minimum) mass of the most massive exoplanet. $^{c}$Semi-major axis of the outermost exoplanet in the system. $^{d}$Eccentricity of innermost exoplanet in the system. $^{e}$\cite{vogt12} report strong evidence for a 5 planet system. A summary of all reported exoplanets in each system is given in Table \ref{summary_table}. Planets for which no eccentricity could be found are assumed to have $e~=~$0.00.}
\tablebib{\textit{Discovery references}: (1) \cite{naef01}; (2) \cite{fischer03}; (3) \cite{udry06}; (4) \cite{butler97}; (5) \cite{butler06}; (6) \cite{tuomi13a}; (7) \cite{queloz00}; (8) \cite{kurster00}; (9) \cite{mayor04}; (10) \cite{pepe11}; (11) \cite{hatzes00}; (12) \cite{galland05}; (13) \cite{jones02}; (14) \cite{mayor09}; (15) \cite{lovis06}; (16) \cite{butler96}; (17) \cite{marcy05}; (18) \cite{tinney11}; (19) \cite{vogt10}; (20) \cite{marcy96}; (21) \cite{dumusque12};  (22) \cite{bonfils05}; (23) \cite{noyes97}; (24) \cite{butler03}; (25) \cite{wright07}; (26) \cite{butler01}; (27) \cite{muterspaugh10}; (28) \cite{cochran97}; (29) \cite{mayor11}; \textbf{(30) \cite{howard11}}; (31) \cite{bailey09}; (32) \cite{marcy99}; (33) \cite{delfosse98}; (34) \cite{mayor95}; (35) \cite{fischer99}.}
\end{table}
\end{landscape}
The appropriate aperture corrections were applied to the flux densities based on the PACS aperture photometry calibration (factors of 0.476, 0.513 and 0.521 for point sources, respectively). Correction factors appropriate to the aperture radius based on the point source encircled energy fraction were applied to the extended sources which, although only an approximation to the true correction, is widely used. A calibration uncertainty of 5\% was assumed for all three PACS bands \citep{balog13}. The flux densities presented in Table \ref{photometry} have not been colour corrected.

The stellar photosphere contribution to the total flux density was calculated from a synthetic stellar atmosphere model interpolated from the PHOENIX/GAIA grid \citep{brott05}. All stars in the sample match the criteria of $d\!<\!25$~pc and luminosity class V except for HD~217107, which is luminosity class IV. The stellar models were scaled to the combined optical, near-infrared and WISE data, where the WISE bands were not saturated or showed evidence of excess emission, following \citet{bertone04}. The stellar physical parameters are given in Table \ref{star_params}. For the DUNES observed targets, the stellar parameters were taken from \cite{eiroa13}. For the DEBRIS observed targets, the stellar parameters were calculated from archival data using the same procedure as for the DUNES targets, see \cite{eiroa13} for details of the method (Maldonado, priv. comm.). Stellar distances were taken from the revised HIPPARCOS catalogue, \citep{vanleeuwen07}. Stellar ages were computed following Eqn. 3 in \cite{mamajek08} for the Ca~{\sc ii} based values whilst the X-ray stellar age is calculated through the relation between X-ray emission and stellar age according to \cite{sanzforcada11}, where most of the ages displayed in Table \ref{star_params} were calculated. Newly calculated X-ray ages are provided for HD~19994, HD~20794, HD~33564, HD~40307, HD~69830, 70~Vir, GJ~581 and HD~192310 as per the method in \cite{sanzforcada11}.
%
\section{Analysis}

A determination of the presence of excess emission from each star was made on the basis of the excess significance, or $\chi$ value. The significance was calculated using the PACS 100 and 160~$\mu$m flux densities and uncertainties in the following manner:

\begin{equation}
  \chi_{\lambda} = \frac{\left(F_{\lambda, \rm obs} - F_{\lambda, \rm pred}\right)}{\sqrt{\sigma^{2}_{\lambda, \rm obs} + \sigma^2_{\lambda, \rm pred} + \sigma^2_{\lambda, \rm cal}}}
\label{eqn_chi}
\end{equation}

Where $F_{\lambda, \rm obs}$ and $F_{\lambda, \rm pred}$ are the observed and predicted (photosphere) flux densities at the wavelength under consideration and the uncertainty is the quadratic sum of the uncertainties of the observation $\sigma_{\lambda, \rm obs}$, the stellar photosphere model $\sigma_{\lambda, \rm pred}$, and calibration $\sigma_{\lambda, \rm cal}$. 

Those stars in the sample with $\chi_{\lambda}\!>\!3$ at either (or both) PACS wavelength(s) were classed as having significant excess emission. The disc fractional luminosity was calculated by fitting a black body emission model parameterised by the dust temperature, $T_{\rm d}$, and fractional luminosity, $L_{\rm IR}/L_{\star}$, to the error-weighted \textit{Spitzer} MIPS 70~$\mu$m and \textit{Herschel} PACS photometry.

For non-excess sources, the dust temperature was assumed to be 37~K, and a 3-$\sigma$ upper limit for the fractional luminosity was then calculated. Adopting a dust temperature of 37~K for calculation provides a strict minimum for the dust fractional luminosity of the non-excess stars based on the observed 100~$\mu$m flux density. The upper limits calculated here (presented in Table \ref{photometry}) are therefore consistent with the typical dust temperature we would expect to observe for debris discs around Sun-like stars.

Although crude, such a model allows a consistent and uniform calculation of the dust properties from the observed flux densities, necessary for the statistical approach to identify trends in the data, without getting distracted by the intricate details of individual sources, e.g. extended emission or material composition.

We note that for disc radii it is well known that an assumption of black body emission can underestimate the radial distance of the dust from the star by a factor of 1--2.5 \citep[for A stars,][]{booth13} or a factor of up to 4 around G and M stars e.g. 61 Vir \citep{wyatt12}, HD~207129 \citep{marshall11,loehne12} and GJ~581 \citep{lestrade12}. We derive radial locations for the dust from the assumption of thermal equilibrium between the dust grains and the incident radiation as a minimum possible orbital distance for the emitting dust in these systems \citep{bp93}.

We calculate values for the basic physical parameters of the discs which are presented in Table \ref{photometry} (dust fractional luminosity, dust temperature and dust radius) from the black body fit to the spectral energy distribution and the assumption of black body absorption and emission for the dust grains. We use the disc fractional luminosity for comparison with the stellar and exoplanet parameters, breaking down the observed sample of exoplanet host stars into three subsamples (low mass planets, cool Jupiters and hot Jupiters) to check for evidence of the trends identified in recent articles \citep[e.g.][]{maldonado12,wyatt12}.

\section{Results}

\begin{figure}
\centering
\includegraphics[width=0.33\textwidth,angle=90]{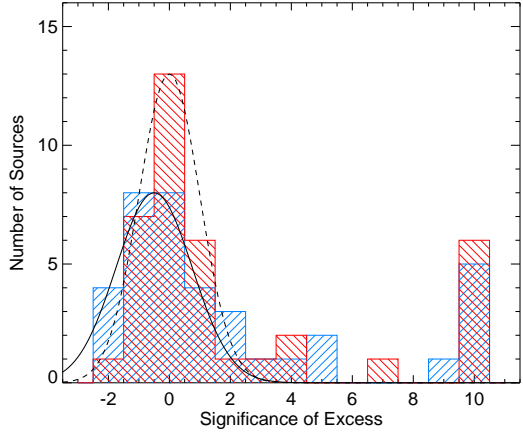}
\caption{Histogram plot of the distribution of the significances presented in Table \ref{photometry} at 100~$\mu$m (in blue, hatching top right to bottom left) and 160~$\mu$m (in red, hatching top left to bottom right). Debris disc stars with $\chi$ values greater than ten have been added to the right hand side bin.}
\label{chi_hist}
\vspace{-0.2in}
\end{figure}

We have detected far-infrared excess emission with $\chi\!>\!3$ from ten of the 37 systems in this sample. HD~69830 is known to have a warm debris disc \citep{beichman05}, bringing the total to 11 systems with both a debris disc and exoplanet(s). The histogram presented in Fig. \ref{chi_hist} illustrates the distribution of the measured (non-)excesses at 100 and 160~$\mu$m, with a long tail toward higher significances, as expected. Fitting Gaussian profiles to both distributions reveals that at 100~$\mu$m the distribution is peaked at $\chi_{100}\!=\!-0.5$ with $\sigma_{100}\!=\!1.25$ whilst at 160~$\mu$m the distribution peakes at $\chi_{160}\!=\!0.0$ with $\sigma_{160}\!=\!1.0$. The peak position and width of the Gaussian are measures of the goodness of our stellar photosphere and uncertainty estimates respectively. We expect (assuming normally distributed uncertainties) that both $\chi$ distributions would peak at $\chi\!=\!0$ with $\sigma\!=\!1$. The discrepancy at 100~$\mu$m could be ascribed to a systematic underestimation of the errors, which is hidden at 160~$\mu$m by the larger uncertainties, but this same trend is seen in the larger DUNES sample of 133 stars \citep{eiroa13}. A physical explanation of the shift of the peak to $\chi_{100}\!<\!0$, implying a deficit in the measured flux density compared to the Rayleigh-Jean extrapolation of the stellar photosphere model from 50~$\mu$m, could be found in the fact that the photosphere models do not take account of the decrease in brightness temperature in the higher layers of the stellar photosphere of Sun-like stars. This decrease in brightness temperature has the effect of reducing the observed flux density below that expected from an extrapolated fit to shorter wavelength measurements, with a greatest effect around 150~$\mu$m \citep{eddy69,avrett03,liseau13}. The magnitude of the deficit is $\sim$20\%, and similar to the magnitude of the measured uncertainty at 100~$\mu$m, around 1 to 2~mJy for these stars. If the magnitude of this effect were to be dependent on stellar spectral type it would also broaden the $\chi$ distribution. The negative bias would be statistically significant should the underlying distribution be Gaussian. Due to the many unknowns in this problem, we adopt the conservative approach of noting that the sample standard deviation is larger than the negative absolute value.

Examining only the FGK stars (three of the 37 stars, one with an excess, are M dwarfs), we obtain a statistical incidence of circumstellar dust of 29.4~$\pm$~9.3~\% (10/34), consistent with the general incidence of 20.2~$\pm$~2.0\% from the DUNES survey \citep{eiroa13}. Previous \textit{Spitzer} results for 45 FGK exoplanet hosts record an incidence of 20.8$^{+7.0}_{-4.6}$~\% \citep{trilling08} and 14.5~$\pm$~3.5~\% 117 FGKM stars by \cite{kospal09}. It should be noted that the detection frequencies quoted for various disc surveys are strongly dependent on the sensitivities, samples and observing strategies adopted in each. Comparable incidences between surveys may therefore be coincidental due to survey differences.

From our analysis of the entire sample, we have detected discs with fractional luminosities in the range 2.4$\times10^{-6}$ to 4.1$\times10^{-4}$ and temperatures from 20~K to 80~K. The 3-$\sigma$ upper limits on the non-detections typically constrain the fractional luminosity to a $\sim$ few $\times10^{-6}$, equivalent to $<~10^{-5}~M_{\oplus}$. The range of dust temperatures is broader than might be expected for typical dust temperatures for debris discs which have the peak of their emission in the far-infrared. In the unusually cold case of HD~210277, the dust temperature can be ascribed to nature of the source, being a candidate 'cold debris disc' \citep{eiroa11}. Four of the circumstellar discs are new detections by \textit{Herschel}: HD~20794, HD~40307, GJ~581 and HD~210277, whilst two more have been confirmed by \textit{Herschel} DUNES after marginal detection by \textit{Spitzer}: HD~19994 and HD~117176. Several of the \textit{Herschel} discovered discs are covered in individual papers, e.g. HD~210277 has been identified as one of the DUNES 'cold debris disc' candidates in \cite{eiroa11}, HD~20794 in \cite{wyatt12} and GJ~581, an M-dwarf debris disc, in \citet{lestrade12}. The final new detection, HD~40307, was noted in the DUNES survey paper \citep{eiroa13}.

\begin{figure*}[!t]
\centering
\subfigure[Ca~{\sc ii} Age vs. fractional luminosity.]{
\includegraphics[width=0.33\textwidth,angle=90]{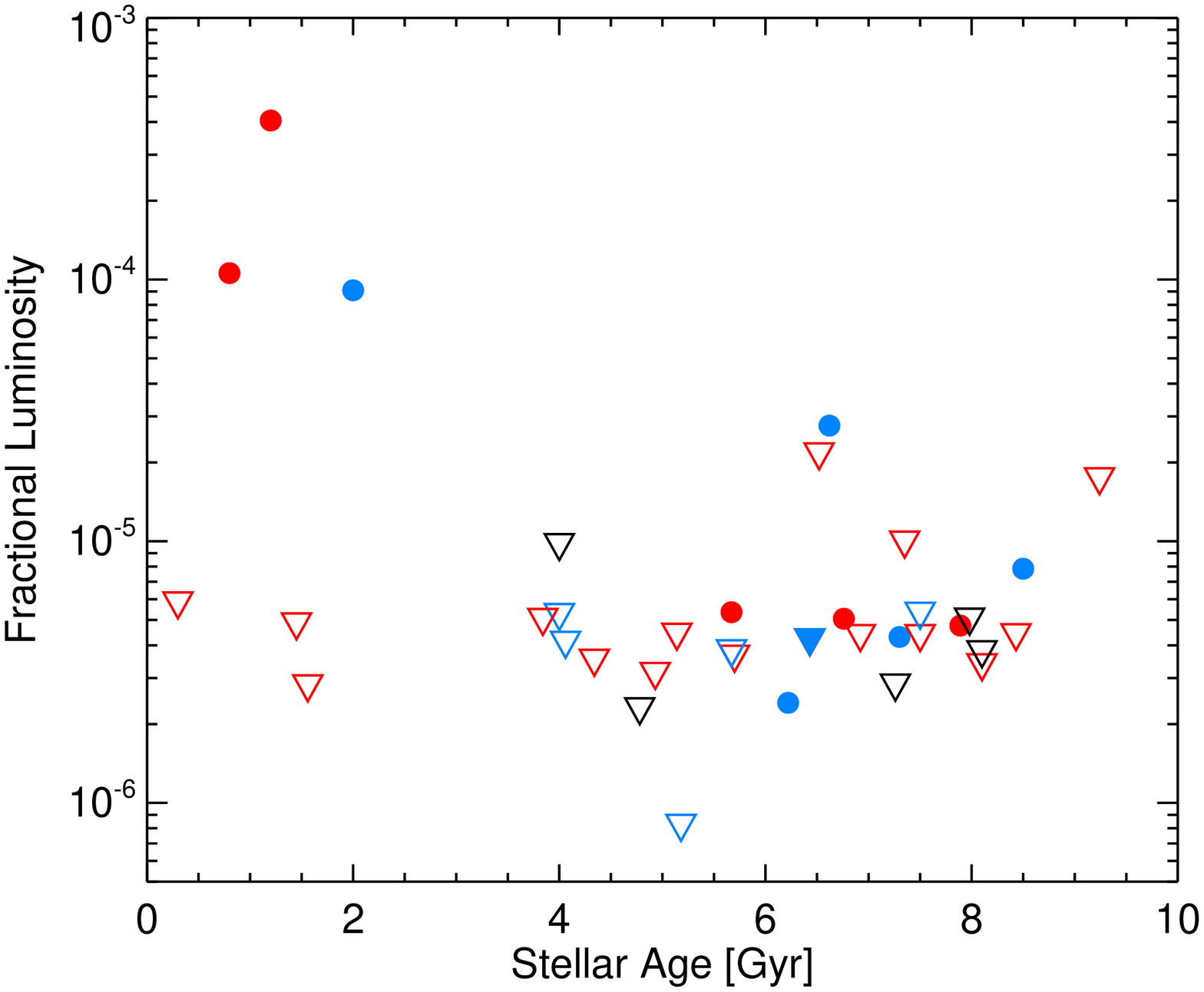}
\label{frac_age}}
\hspace{0.1in}
\subfigure[Metallicity vs. fractional luminosity.]{
\includegraphics[width=0.33\textwidth,angle=90]{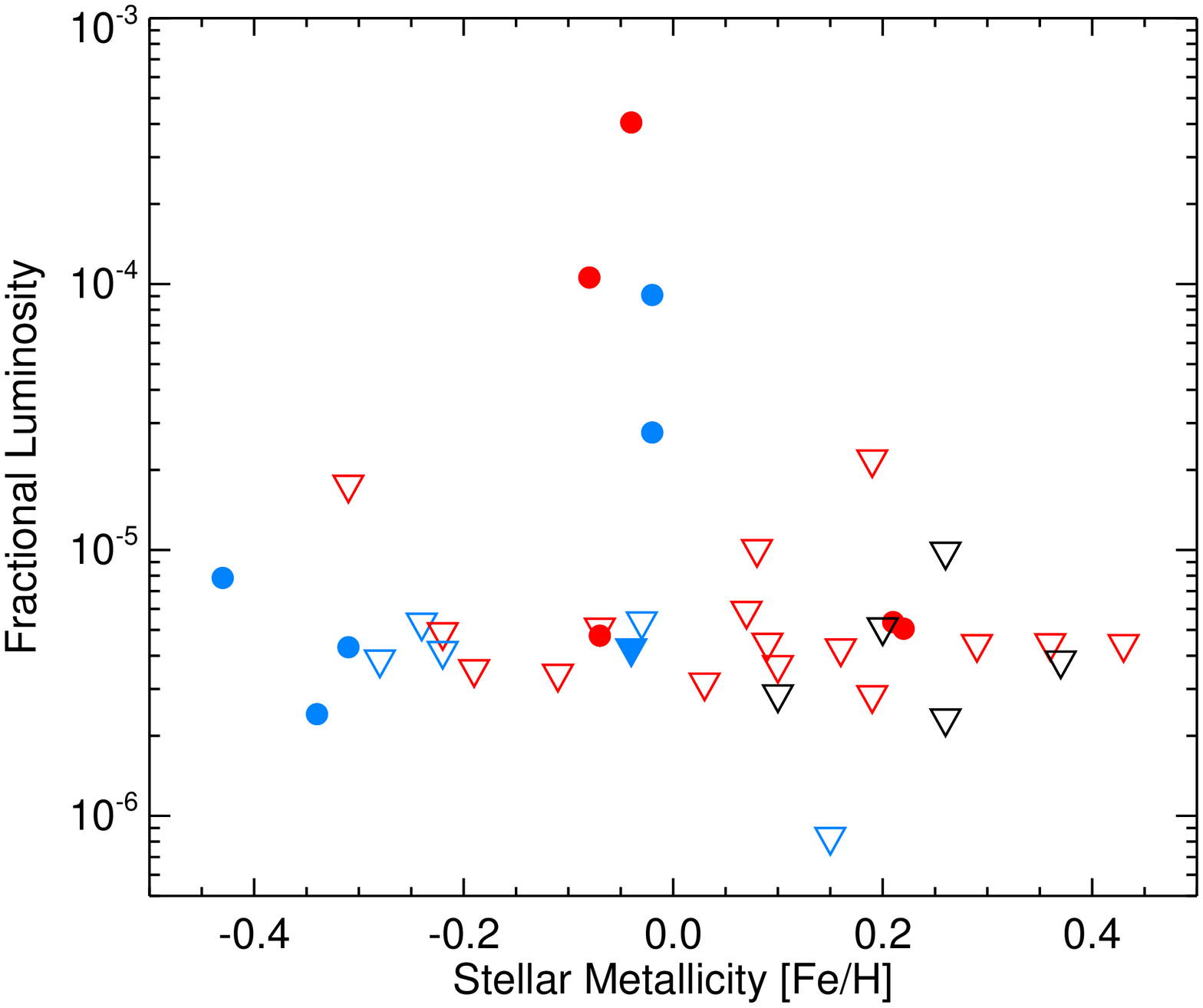}
\label{frac_metal}}
\vspace{0.1in}
\subfigure[Photospheric temperature vs. fractional luminosity.]{
\includegraphics[width=0.33\textwidth,angle=90]{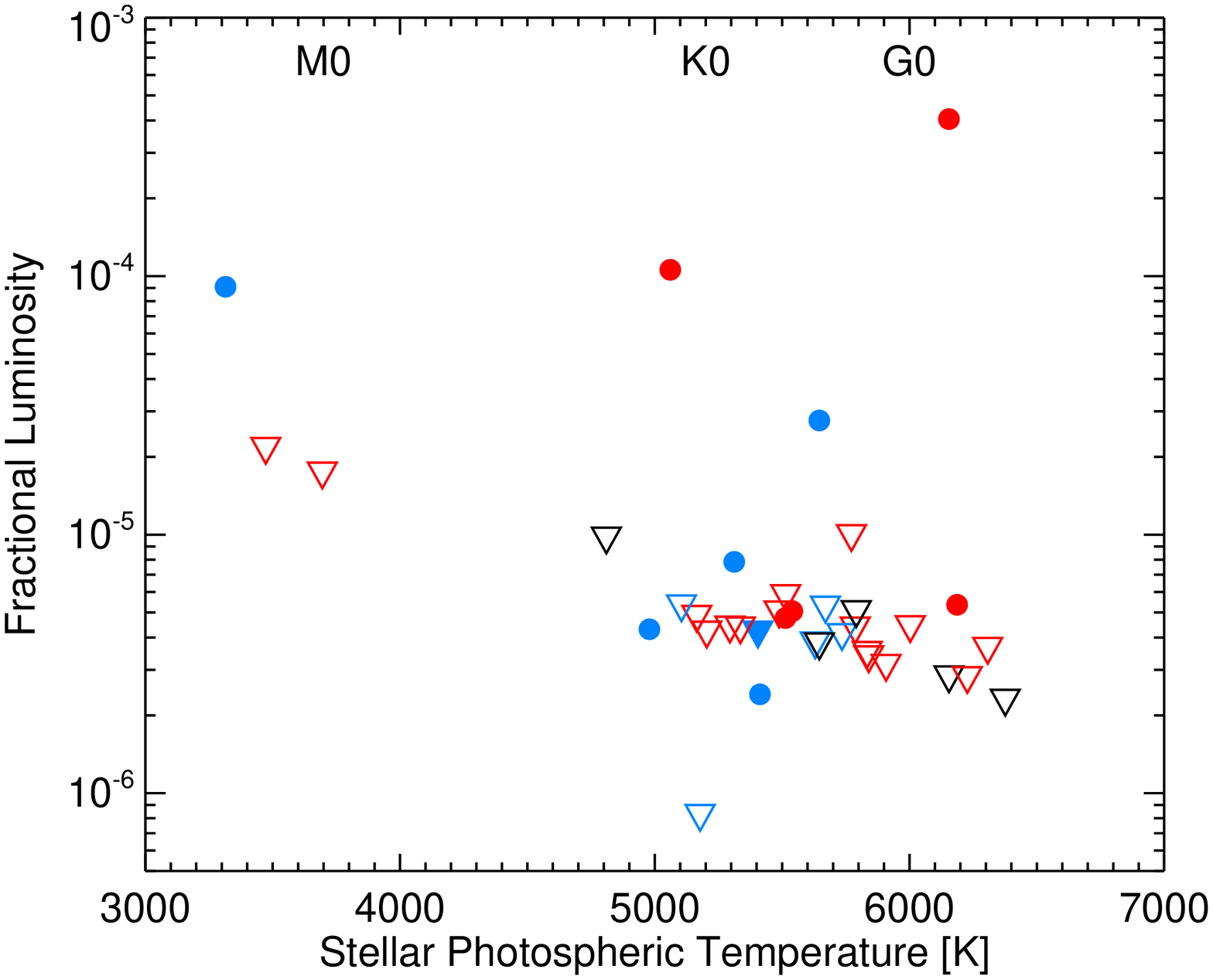}
\label{frac_temp}}
\hspace{0.1in}
\subfigure[Mass of most massive exoplanet vs. fractional luminosity.]{
\includegraphics[width=0.33\textwidth,angle=90]{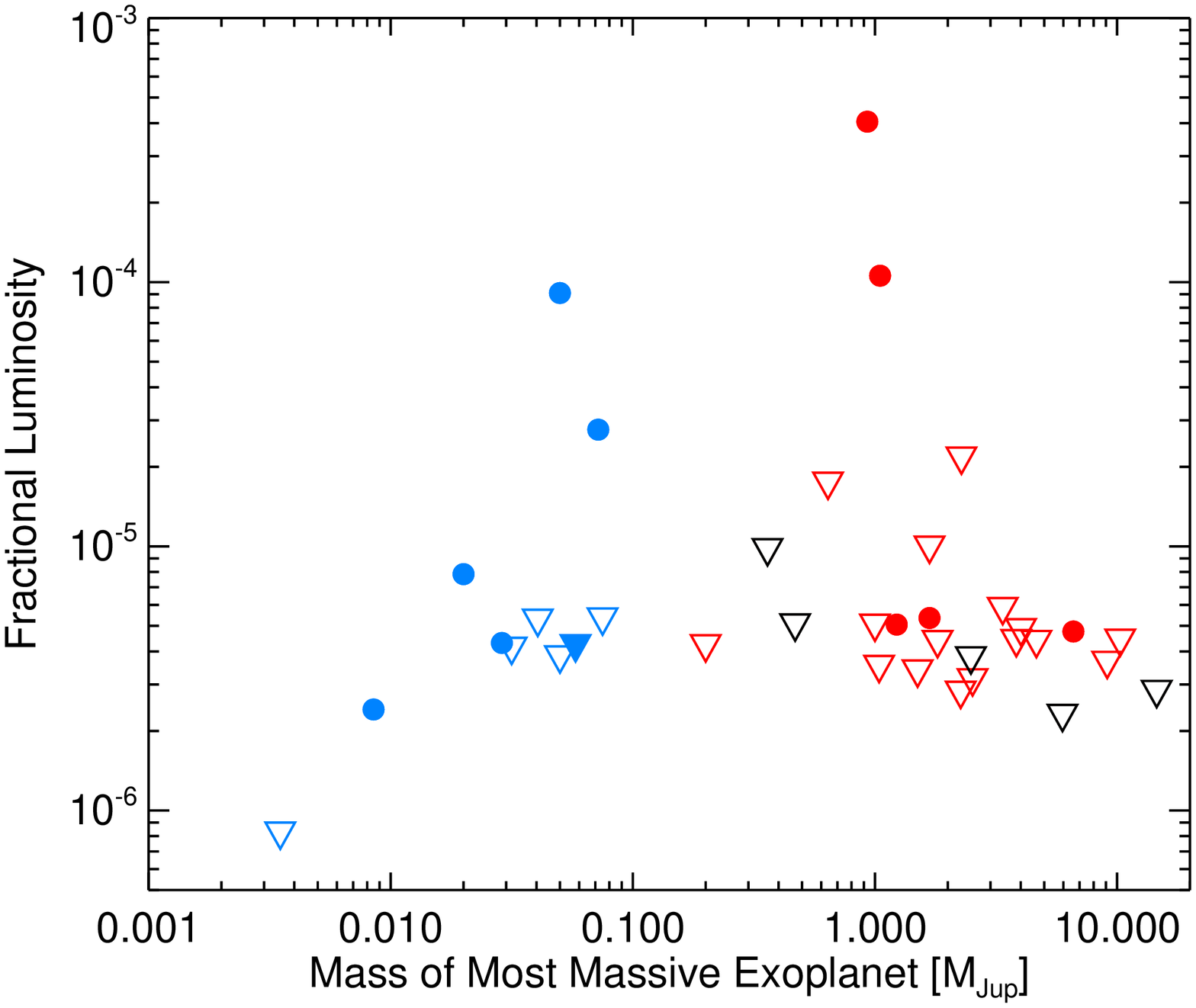}
\label{frac_mplt}}
\vspace{0.1in}
\subfigure[Orbital eccentricity of the innermost exoplanet vs. fractional luminosity.]{
\includegraphics[width=0.33\textwidth,angle=90]{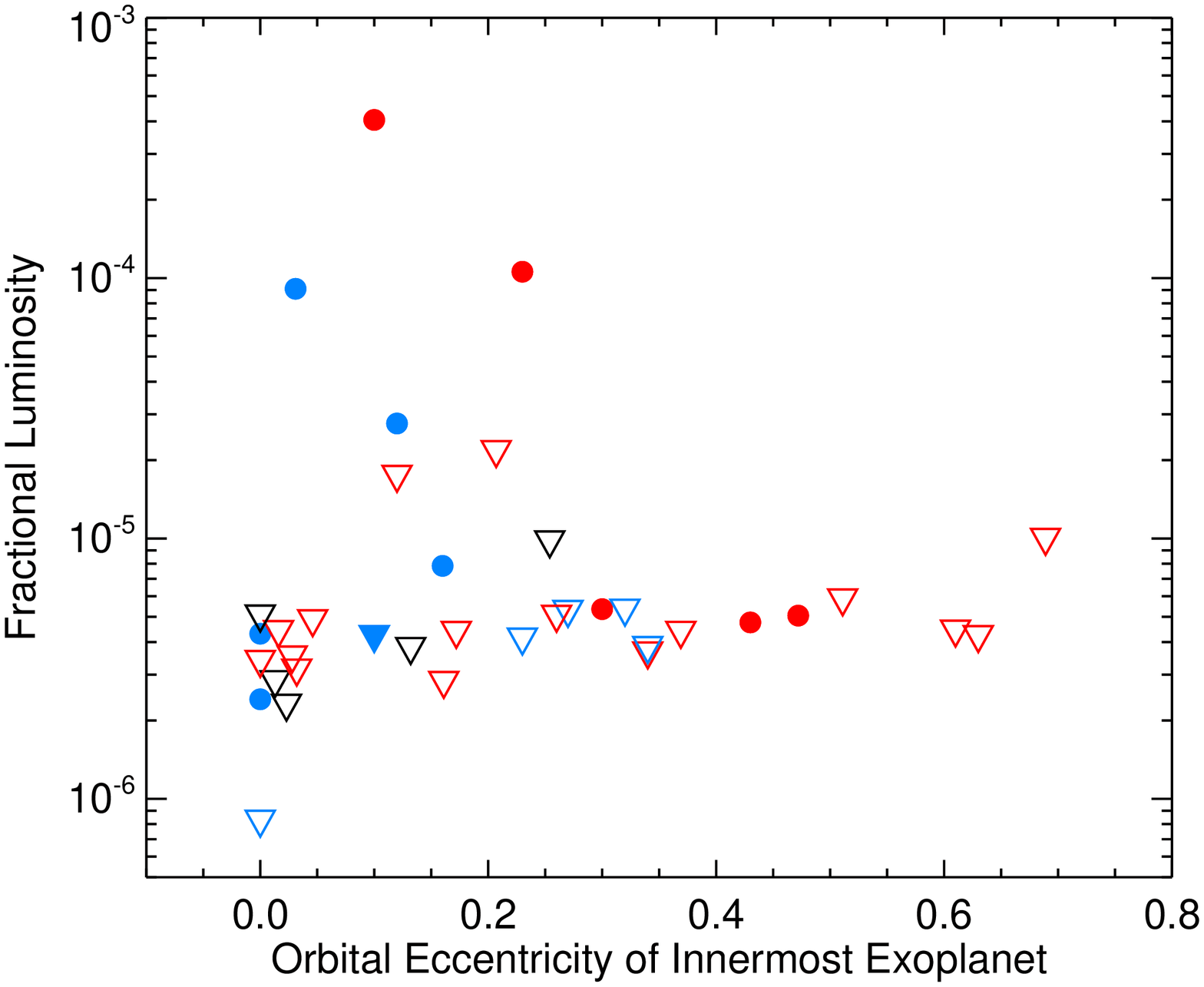}
\label{frac_eplt}}
\hspace{0.1in}
\subfigure[Semi-major axis of outer exoplanet's orbit vs. fractional luminosity.]{
\includegraphics[width=0.33\textwidth,angle=90]{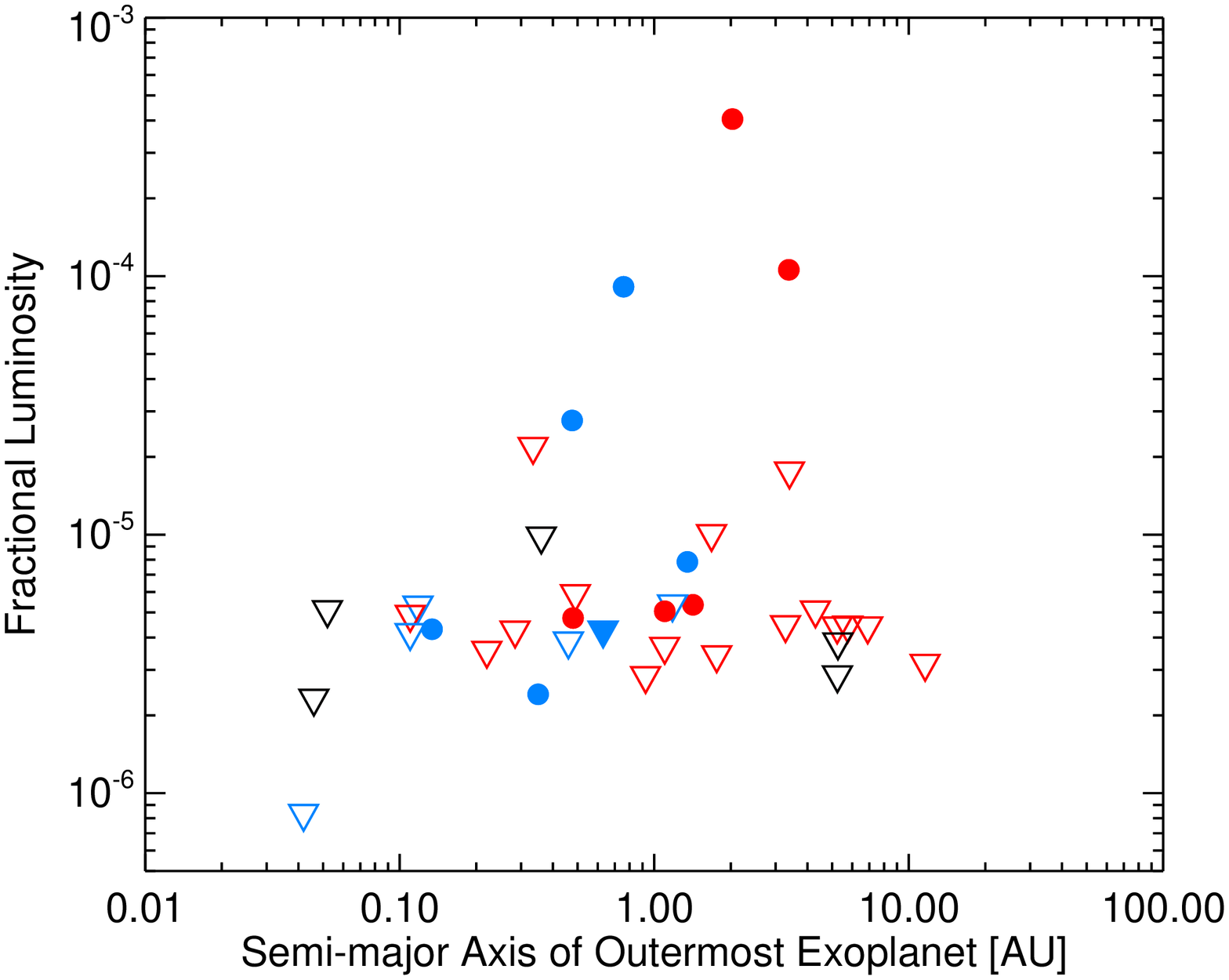}
\label{frac_rplt}}
\caption{Filled circles denote cool debris disc systems whilst open triangles denote fits to the 3-$\sigma$ upper limits at 100~$\mu$m. HD~69830, having a warm debris disc, is denoted as a filled blue triangle to distinguish it from the other far-infrared non-detections. Red data points are cool giant planet systems, black data points are hot giant planet systems whilst blue data points are low mass planet systems.}
\label{results_fig}
\end{figure*}

\section{Discussion}

\subsection{Contamination}

A critical issue with the attribution of an infrared excess to circumstellar dust is the chance of contamination by alignment along the line of sight with a background source. 

The likelihood that one (or more) of the debris discs in the sample have been spuriously identified as such was calculated from the contamination probabilities within both a radius equivalent to one beam HWHM and the positional offset between the expected and observed source position of all the debris disc stars. The source number counts in DEBRIS survey fields from \citet{sibthorpe12} (their equation Eqn. 2) were used to quantify the background source number density at 100~$\mu$m and 160~$\mu$m for sources with flux densities equal to or brighter than the observed excesses of the debris disc stars presented here. Firstly, the probability of confusion for each individual source, $n$, was calculated, including the specific SNR and flux density, giving $P_{n, \rm{conf}}$. Secondly, the probability that none of the sources were confused, $P_{\rm none}$, was calculated by multiplication of the probabilities for the individual sources, i.e. $P_{\rm none}\!=\!P_{1,\rm{conf}} \times P_{2,\rm{conf}} \times \ldots \times P_{n,\rm{conf}}$. The probability of confusion for the whole sample is then simply $1\!-\!P_{\rm none}$.

Within one beam HWHM, the probability of contamination of at least one source amongst the 37 is 1.4\% at 100~$\mu$m and 8.2\% at 160~$\mu$m. Within the maximum position offset (6.7\arcsec), the probability of confusion increases to 4.8\% at 100~$\mu$m and 11.0\% at 160~$\mu$m. In the worst case, based on the maximum offset radius, two stars at 100~$\mu$m and four stars at 160~$\mu$m out of the 37 could be the result of contamination. 

\subsection{Correlations}

We have plotted the calculated dust fractional luminosities (or upper limits) of the exoplanet stars against several physical parameters of the host stars and exoplanets in order to look for trends in the sample. For Fig. \ref{results_fig}(d)--(f) we have adopted the same exoplanet parameters to characterise each system as used in \cite{maldonado12} so a comparison between our findings and theirs can be made more easily. We categorise the type of exoplanet system (low mass, hot Jupiter or cold Jupiter) according to the planet mass where $M_{\rm Planet}\!>\!30~M_{\oplus}$ is a giant planet and a system with a planet of $R_{\rm Planet}\!<\!0.1~AU$ and $M_{\rm Planet}\!>\!30~M_{\oplus}$ defines a hot Jupiter system. Within the sample there are 11 low mass planet systems, five hot Jupiter systems and 21 cold Jupiter systems. Although no cold dust emission from HD~69830's circumstellar disc has been detected it is included as a debris disc star in the statistics but not marked as an excess source on the plots in Fig. \ref{results_fig}, which were based only on far-infrared emission. To quantify the significance of any trend observed in the figures we use the Fisher exact probability test, which has the virtue (compared to a $\chi^{2}$ test) of producing meaningful results even for low sample sizes ($N\!<\!5$), however, the test only gives a probability in support of the null hypothesis (i.e. that both the samples are from the same underlying distribution). For example, comparing the number of debris disc stars around the FGK exoplanet host stars in this sample (34 stars, 10 discs) to the \textit{Spitzer} and \textit{Herschel} DUNES frequencies \citep{trilling08,eiroa13} we find that the resultant $p$-values are 0.19 and 0.34, respectively. The exoplanet host sample is not significantly different from these larger samples and can be presumed to have been drawn from the same underlying population, despite the inherent biases of being composed solely of exoplanet host stars and the difference in sensitivity of the \textit{Spitzer} MIPS and \textit{Herschel} PACS based surveys.

The relationship between stellar age and fractional luminosity for circumstellar discs, presented in Fig. \ref{frac_age}, has been explored for a broad range of ages by e.g. \cite{decin03}, \cite{hernandez07} and \cite{wyatt07}. In this work we have used stellar ages derived from Ca~{\sc ii} H \& K activity and X-ray luminosity taken from \cite{eiroa13} for the DUNES observed sources or calculated in the same manner as the DUNES sources but based on publicly available data (Maldonado and Sanz Forcada, priv. comm.). Binning the sample into three broad age ranges we find that the incidence of dust weakly decreases with age, albeit with error bars large enough that the incidence could be constant across the bins considered here, from 50~$\pm$~29\% ($t\!<$~3.5~Gyr) to 28~$\pm$~12\% ($t\!=\!$~3.5--7.0~Gyr) to 25~$\pm$~14\% ($t\!>$7.0~Gyr). The decay of debris discs around Sun-like stars is examined in greater detail in \cite{kains11}.

The sample of exoplanet host stars is almost evenly divided between sub- and supra-Solar metallicity, with 18 and 19 stars in each subsample, which is shown in Fig \ref{frac_metal}. Of the stars with greater metallicities than the Sun, five host a hot Jupiter planet, whilst none of the 18 low metallicity stars have hot Jupiters, giving a $p$-value of 0.04, i.e. the hot Jupiters and other planets do not come from the same underlying distribution, therefore we are able to identify the well known correlation between increased stellar metallicity and hot Jupiters from our data set \citep{fv05,greaves06,maldonado12}. Comparing the distribution of hot and cold Jupiters versus metallicity we find a $p$-value of 0.58, and therefore we cannot differentiate between the stars with wide and close orbiting Jovian exoplanets from this sample.

All the low mass planetary systems (i.e. those where the most massive planet is $<~30~M_{\oplus}$) except $\alpha$ Cen B have host stars with sub-Solar metallicity (10/18), consistent with \cite{jenkins12}. Stars hosting low mass planets have not been found to be preferentially metal rich, unlike stars hosting high mass planets \citep{santos01,ghezzi10,buchhave11,mayor11}. Comparing the incidence of low mass and high mass planets around stars with sub- or supra-Solar metallicity, we obtain a probability $p$-value of 0.001, suggesting that both the low-mass and high-mass planet stars are not drawn from the same underlying population. This correlation between low mass planets and low metallicity stars could be the product of two phenomena: in a low metallicity disc there will be fewer solids from which to form massive planetary cores that go on to become gas giant planets, making such planets rarer around those stars \citep{greaves07,wyatt07}; additionally, protoplanetary discs with lower metallicity are thought to be dispersed more easily and quickly, because of the lower optical depth, so UV and X-rays penetrate further into the disc and cause the gas loss through winds to be stronger, such that any nascent giant planet must capture its gaseous envelope more quickly as the disc dispersal occurs more rapidly in such systems, thereby limiting the number of gas giants that will form \citep{yasui09,ercolano10}.

We find an incidence of debris around 10/18 stars in the low metallicity group, and 2/19 in the high metallicity group. This implies an anti-correlation between the presence of debris and metallicity, with a $p$-value of 0.005. The presence of brighter debris discs around low metallicity stars could be inferred to represent the inability of the low mass planets to scatter the dust producing planetesimals near their formation region, or via migration from a larger initial semi-major axis, as effectively as a gas giant, as proposed in \cite{wyatt12}. The observed anticorrelation may not be therefore directly related to the properties of the debris disc and stellar metallicity, but could be a consequence of the type of planets that form around low metallicity stars as stated above and can be explained within the context of a model in which planets form through core accretion.

The effect of the stellar temperature on the determination of the upper limit to the dust fractional luminosity can be seen in Fig. \ref{frac_temp}. For the coolest stars in the sample, we are limited to disc brightnesses of $\geq\!2\times10^{-5}$, whilst at the hotter end we can put better constraints of $\geq\!2\times10^{-6}$ on the dust present around those stars. For most stars in our sample we have obtained 3-$\sigma$ upper limits on the fractional luminosity of a few $\times10^{-6}$, i.e. a few ten times greater than that expected of the EKB \citep{vitense12}. There is no visible correlation between the stellar photosphere temperature and the presence of dust around the star. Dividing the sample by spectral type, we find debris discs around 33.3~$\pm$~23.6~\% of F stars (2/6), 27.3~$\pm$~11.1~\% of G stars (6/22), 33.3~$\pm$~23.6~\% of K stars (2/6) and 33.3~$\pm$~33.3\% of M stars (1/3). These values are consistent with the measurements from surveys with much larger samples, e.g. \cite[][]{trilling08,kospal09,eiroa13} for FGK stars and \cite{lestrade06,lestrade09,gautier07} for M stars.

As seen in Fig. \ref{frac_mplt}, the exoplanets around the \textit{Herschel} discovered debris disc stars all have masses  $M_{\rm Planet}\!<\!30~M_{\oplus}$. Comparing the most massive exoplanet in each system to the dust fractional luminosity reveals that the stars with brighter discs generally have a low mass planet ($M_{\rm Planet}\!<\!30~M_{\oplus}$), being 6/11 stars for the low mass subset c.f. 5/26 for the remainder of the sample. The $p$-value for the comparison of these two sub-samples is 0.05. Stars with only low mass planets are more likely to harbour debris discs than those stars with Jovian planet(s), consistent with the findings of \cite{maldonado12} and also confirming the trend suggested in \cite{wyatt12} (based on a smaller sample of G stars).

The relationship between eccentricity, characterised by the orbital eccentricity of the innermost exoplanet, and fractional luminosity illustrated in Fig. \ref{frac_eplt} appears to favour the presence of debris discs around stars with low eccentricity planetary systems ($e\!<\!0.2$). Splitting the sample at $e\!=\!0.2$, we find a greater incidence of dust in low eccentricity systems (7/20) over high eccentricity systems (4/17), agreeing with the predictions (for giant planets) in \cite{raymond11} -- see their Figs. 13 and 18, and similarly consistent with the results of \cite{maldonado12}. However, putting these sub-samples to the test we find that the $p$-value is 0.49, which is an inconclusive result, but suggestive that the two groups are more similar than not, which may be interpreted as illustrating the known exoplanets in these systems have little dynamical influence on the visible debris.

Similarly, a comparison of the fractional luminosities of systems with hot Jupiters and those with cold Jupiters in Fig. \ref{frac_rplt} looks like the cold Jupiter systems have a greater tendency to host a debris disc. We see that none of the five hot Jupiters host stars subset are observed to have any excess emission, whilst five of the 21 cold Jupiter planet host stars do have a detectable debris disc. In this case the Fischer test again returns an intermediate result, with $p$-value is $0.54$. We cannot therefore confirm the trend identified in \cite{maldonado12} between cool giant planets and fainter debris discs based on the observations analysed here.

%
\section{Conclusions}

We have presented \textit{Herschel} PACS observations of 37 nearby exoplanet systems from the DUNES and DEBRIS samples aimed at searching for exo-EKB analogues. Excess emission attributable to the presence of a circumstellar debris disc was observed around ten of these stars; for the remaining stars we found no evidence of significant excess emission, including the non-measurement of cold emission around HD~69830, providing a tighter upper limit to any possible cold dust in that system. We have improved on the upper limits for dust detection for the stars in this sample by a factor of two over previous \textit{Spitzer} observations, constraining the possible flux from cold dust in all observed systems to at worst two orders of magnitude greater that of the EKB and in several cases at levels comparable to that of the EKB.

We find incidences of $\sim$~30~\% for cool debris discs around exoplanet host stars from the sample examined here, irrespective of the spectral type. Due to the large uncertainties in this measurement (from the small sample sizes), these values are in fact consistent with the incidence of debris discs measured by DUNES 20.2~$\pm$~2.0~\% \citep{eiroa13}. The incidence of debris is seen to decrease around older stars, again with large uncertainties.

We have identified several trends between the stellar metallicity, the presence of a debris disc and the mass of the most massive exoplanet around the star. We found that low metallicity stars are more likely to host low mass planets, low metallicity stars are also more likely to have a detectable debris disc and that low mass planets are more likely to be associated with a detectable debris disc. This is consistent with what would be expected from the core accretion planet formation model. Combining these trends we develop a picture for these systems in which the gas is stripped from the protoplanetary disc too quickly for Jovian mass planets to form and the resulting low-mass planets can't scatter planetesimals as strongly as more massive planets if they form in-situ, or as they migrate through a disc from a larger initial semi-major axis \citep{maldonado12,wyatt12}.

Furthermore, we find no significant evidence for a trend relating the eccentricity of the innermost planet with the fractional luminosity, suggesting that the known exoplanets in these systems have little influence on the visible presence of dust, which is unsurprising as the two components are well separated. We also find no evidence to support the proposed trend between cold Jupiters and lower dust luminosities proposed in \cite{maldonado12}, though this is unsurprising since the sample analysed here is only a sub-set of those from Maldonado et al's work and the newly discovered \textit{Herschel} debris discs have been found exclusively around low mass planet host stars.

As an extention to this analysis a companion paper to this work is in preparation, comparing the exoplanets sample presented here to unbiased control samples of stars with exoplanets, or a debris disc, or without either. This future paper will look for differences in the incidence of dusty debris between these different sub-samples. In future work, observations from the \textit{Herschel} SKARPS Open Time Programme \citep[Search for Kuiper Belts Around Radial-Velocity Planet Stars;][]{kennedy13} will be used to increase the number of radial velocity planet host stars for which far-infrared fluxes are available, clarifying, and hopefully supporting, the trends seen between dust, planet and stellar properties in this, and earlier, works.
%
\bibliographystyle{aa}
\bibliography{dunes_debris_exoplanets}
%
\begin{acknowledgements}
This research has made use of NASA's Astrophysics Data System Bibliographic Services. This research has made use of the SIMBAD database, operated at CDS, Strasbourg, France. This research has made use of the Exoplanet Orbit Database and the Exoplanet Data Explorer at exoplanets.org, and the Extrasolar Planets Encyclopedia at exoplanets.eu. JPM, CE, JM and BM are partially supported by Spanish grant AYA 2011-26202. This work was supported by the European Union through ERC grant number no. 279973 (GMK and MCW) and has been partially supported by \textit{Spitzer} grant OT1\_amoromar\_1 (AMM).
\end{acknowledgements}
%

\onecolumn
\begin{longtable}{llccccc}
\caption{Summary table of all exoplanets in each system.} \label{summary_table}\\
\hline
\hline
Name & Planet & Mass ($M\sin i$) & Semi-major axis  & Eccentricity &\multicolumn{2}{c}{Reference}\\
       &        & [$M_{\rm Jup}$]  & [AU]           &              & Orbit & Discovery \\
\hline
\endfirsthead
\caption{Continued.} \\
\hline\hline
Name & Planet & Mass ($M\sin i$) & Semi-major axis  & Eccentricity &\multicolumn{2}{c}{Reference}\\
       &        & [$M_{\rm Jup}$]  & [AU]           &              & Orbit & Discovery \\
\hline
\endhead
\hline
\endfoot
\hline
\endlastfoot
HD~1237 & b & 3.37~$\pm$~0.09 & 0.4947~$\pm$~0.0083 & 0.511~$\pm$~0.017 & 1 & 1 \\
HD~3651 & b & 0.229~$\pm$~0.0148 & 0.2947~$\pm$~0.0049 & 0.596~$\pm$~0.036 & 2 & 3 \\
HD~4308 & b & 0.0477~$\pm$~0.0028 & 0.1192~$\pm$~0.00199 & 0.000~$\pm~^{0.01}_{-0.00}$ & 4 & 4 \\
$\upsilon$~And & b & 0.669~$\pm$~0.026 & 0.05939~$\pm$~0.00099 & 0.013~$\pm~^{0.016}_{0.013}$ & 5 & 6 \\
         & c & 1.919~$\pm$~0.088 & 0.830~$\pm$~0.0138 & 0.224~$\pm$~0.026 & 5 & 7 \\
         & d & 4.12~$\pm$~0.165 & 2.525~$\pm$~0.042 & 0.267~$\pm$~0.0196 & 5 & 7 \\
q$^{1}$~Eri & b & 0.93~$\pm$~0.24 & 2.022~$\pm$~0.082 & 0.16~$\pm~^{0.22}_{0.16}$ & 8 & 8 \\
$\tau$~Ceti & b & 0.0063~$\pm$~0.0025 & 0.105~$\pm$~0.005 & 0.16~$\pm^{0.22}_{0.16}$ & 9 & 9 \\
         & c & 0.0098~$\pm^{0.0044}_{0.0035}$ & 0.195~$\pm$~0.009 & 0.03~$\pm^{0.28}_{0.03}$ & 9 & 9 \\
         & d & 0.0113~$\pm$~0.0054 & 0.374~$\pm$~0.020 & 0.08~$\pm^{0.26}_{0.08}$ & 9 & 9 \\
         & e & 0.0135~$\pm$~0.0066 & 0.552~$\pm$~0.023 & 0.05~$\pm^{0.22}_{0.05}$ & 9 & 9 \\
         & f & 0.0208~$\pm$~0.0110 & 1.350~$\pm$~0.080 & 0.03~$\pm^{0.26}_{0.03}$ & 9 & 9 \\
GJ~86 & b & 4.00~$\pm$~0.137 & 0.1142~$\pm$~0.0019 & 0.0416~$\pm$~0.0072 & 8 & 10 \\
$\iota$~Hor & b & 2.05~$\pm$~0.2 & 0.924~$\pm$~0.0161 & 0.14~$\pm$~0.13 & 8 & 11 \\
HD~19994 & b & 1.33~$\pm$~0.105 & 1.306~$\pm$~0.022 & 0.266~$\pm$~0.014 & 2 & 12 \\
HD~20794 & b & 0.00849~$\pm$~0.00096 & 0.1207~$\pm$~0.002 & 0 & 13 & 13 \\
          & c & 0.0074~$\pm$~0.00135 & 0.2036~$\pm$~0.0034 & 0 & 13 & 13 \\
          & d & 0.0148~$\pm$~0.00181 & 0.3498~$\pm$~0.0058 & 0 & 13 & 13 \\
$\epsilon$~Eri & b & 1.05~$\pm$~0.188 & 3.38~$\pm$~0.32 & 0.25~$\pm$~0.23 & \textbf{8} & \textbf{14} \\
HD~33564 & b & 9.13~$\pm$~0.37 & 1.124~$\pm$~0.0196 & 0.340~$\pm$~0.02 & 15 & 15 \\
HD~39091 & b & 10.09~$\pm$~0.38 & 3.35~$\pm$~0.104 & 0.6405~$\pm$~0.0072 & 8 & 16 \\
HD~40307 & b & 0.01291~$\pm$~0.00084 & 0.04689~$\pm$~0.00078 & 0 & 17 & 17 \\
          & c & 0.0211~$\pm$~0.00118 & 0.0801~$\pm$~0.00133 & 0 & 17 & 17 \\
          & d & 0.0281~$\pm$~0.00162 & 0.1324~$\pm$~0.0022 & 0 & 17 & 17 \\
HD~69830 & b & 0.0316~$\pm$~0.00172 & 0.0782~$\pm$~0.0013 & 0.100~$\pm$~0.04 & 18 & 18 \\
          & c & 0.0368~$\pm$~0.0025 & 0.1851~$\pm$~0.0031 & 0.130~$\pm$~0.06 & 18 & 18 \\
          & d & 0.0563~$\pm$~0.0052 & 0.627~$\pm$~0.0122 & 0.070~$\pm$~0.07 & 18 & 18 \\
55~Cnc & b & 0.801~$\pm$~0.027 & 0.1134~$\pm$~0.00189 & 0.0040~$\pm$~0.003 & \textbf{19} & 6 \\
          & c & 0.1646~$\pm$~0.0066 & 0.2373~$\pm$~0.004 & 0.070~$\pm$~0.02 & 19 & 20 \\
          & d & 3.54~$\pm$~0.122 & 5.475~$\pm$~0.094 & 0.0200~$\pm$~0.008 & 19 & 20 \\
          & e & 0.0262~$\pm$~0.00123 & 0.01544~$\pm$~0.00026 & 0 & 19 & \textbf{21} \\
          & f & 0.173~$\pm$~0.0106 & 0.774~$\pm$~0.0129 & 0.320~$\pm$~0.05 & 19 & \textbf{22} \\
47 UMa & b & 2.546~$\pm$~0.096 & 2.101~$\pm$~0.035 & 0.032~$\pm$~0.014 & 23 & 24 \\
          & c & 0.546~$\pm$~0.071 & 3.57~$\pm$~0.111 & 0.098~$\pm$~0.071 & 23 & 25 \\
HD~99492 & b & 0.106~$\pm$~0.0117 & 0.1219~$\pm$~0.002 & 0.254~$\pm$~0.092 & 8 & 26 \\
          & c & 0.36~$\pm$~0.02 & 5.4~$\pm$~0.1 & 0.106~$\pm$~0.006 & 27 & 27 \\
HD~102365 & b & 0.0510~$\pm$~0.0081 & 0.4633~$\pm$~0.0078 & 0.34~$\pm$~0.14 & 28 & 28 \\
61~Vir & b & 0.0161~$\pm$~0.00184 & 0.05006~$\pm$~0.00083 & 0.12~$\pm$~0.11 & 29 & 29 \\
          & c & 0.0334~$\pm$~0.0038 & 0.2169~$\pm$~0.0036 & 0.140~$\pm$~0.06 & 29 & 29 \\
          & d & 0.0716~$\pm$~0.0093 & 0.4745~$\pm$~0.008 & 0.350~$\pm$~0.09 & 29 & 29 \\
70~Vir & b & 7.46~$\pm$~0.25 & 0.4836~$\pm$~0.0081 & 0.4007~$\pm$~0.0035 & 8 & 30 \\
$\tau$~Boo & b & 4.17~$\pm$~0.142 & 0.04800~$\pm$~0.0008 & 0.023~$\pm$~0.015 & 31 & 6 \\
$\alpha$~Cen B & b & 0.00356~$\pm$~0.00028 & 0.0419~$\pm$~0.0007 & 0 & 32 & 32 \\
GJ~581 & b & 0.0499~$\pm$~0.0023 & 0.04061~$\pm$~0.00087 & 0.031~$\pm$~0.014 & 33 & 34 \\
          & c & 0.0168~$\pm$~0.00119 & 0.0729~$\pm$~0.00157 & 0.070~$\pm$~0.06 & 33 & 35 \\
          & d & 0.0191~$\pm$~0.0022 & 0.2177~$\pm$~0.0047 & 0.250~$\pm$~0.09 & 33 & 35 \\
          & e & 0.00613 ~$\pm$~0.00071 & 0.02846~$\pm$~0.00061 & 0.320~$\pm$~0.09 & 33 & 17 \\
$\rho$~CrB & b & 1.064~$\pm$~0.053 & 0.2257~$\pm$~0.0038 & 0.057~$\pm$~0.028 & 8 & 36 \\
14~Her & b & 5.21~$\pm$~0.3 & 2.934~$\pm$~0.084 & 0.3690~$\pm$~0.005 & 37 & 38 \\
HD~154345 & b & 0.957~$\pm$~0.061 & 4.21~$\pm$~0.105 & 0.044~$\pm~^{0.046}_{0.044}$ & 39 & 40 \\
$\mu$~Ara & b & 1.746~$\pm$~0.069 & 1.527~$\pm$~0.029 & 0.128~$\pm$~0.017 & 41 & 42 \\
          & c & 1.89~$\pm$~0.22 & 5.34~$\pm$~0.4 & 0.099~$\pm$~0.063 & 41 & 43 \\
          & d & 0.0346~$\pm$~0.00199 & 0.0928~$\pm$~0.00177 & 0.172~$\pm$~0.04 & 41 & 44 \\
          & e & 0.543~$\pm$~0.03 & 0.940~$\pm$~0.018 & 0.067~$\pm$~0.0122 & 41 & 41,45 \\
HD~176051 & b & 1.5~$\pm$~0.3 & 19.1 & 0.2667~$\pm$~0.0022 & 46 & 46 \\
16~Cyg~B & b & 1.640~$\pm$~0.083 & 1.660~$\pm$~0.028 & 0.681~$\pm$~0.017 & 8 & 47 \\
HD~189567 & b &  0.0316~$\pm$~0.0034 & 0.1099~$\pm$~0.0018 & 0.23~$\pm$~0.14 & 48 & 48 \\
HD~192310 & b & 0.0531~$\pm$~0.0028 & 0.3223~$\pm$~0.0054 & 0.130~$\pm$~0.04 & \textbf{13} & 49 \\
          & c & 0.0736~$\pm$~0.0098 & 1.184~$\pm$~0.024 & 0.32~$\pm$~0.11 & \textbf{13} & \textbf{13} \\
GJ~832 & b & 0.644~$\pm$~0.075 & 3.40~$\pm$~0.153 & 0.12~$\pm$~0.11 & 50 & 50 \\
HD~210277 & b & 1.273~$\pm$~0.051 & 1.131~$\pm$~0.0189 & 0.476~$\pm$~0.017 & 8 & 51 \\
GJ~876 & b & 1.95~$\pm$~0.122 & 0.2081~$\pm$~0.0065 & 0.0324~$\pm$~0.0013 & 52 & 53,54 \\
           & c & 0.612~$\pm$~0.021 & 0.1296~$\pm$~0.0022 & 0.25591~$\pm$~0.00093 & 52 & 55 \\
           & d & 0.0184~$\pm$~0.00123 & 0.02081~$\pm$~0.00035 & 0.207~$\pm$~0.055 & 52 & 56 \\
           & e & 0.0392~$\pm$~0.0051 & 0.333~$\pm$~0.0105 & 0.055~$\pm$~0.012 & 52 & 52 \\
51~Peg & b & 0.461~$\pm$~0.0164 & 0.05211~$\pm$~0.00087 & 0.013~$\pm$~0.012 & 8 & 57 \\
HD~217107 & b & 1.401~$\pm$~0.048 & 0.0750~$\pm$~0.00125 & 0.1267~$\pm$~0.0052 & 5 & 58 \\
           & c & 2.62~$\pm$~0.15 & 5.33~$\pm$~0.2 & 0.517~$\pm$~0.033 & 5 & 59 \\
\end{longtable}
\textit{References}: (1) \cite{naef01}; (2) \cite{wittenmyer09}; (3) \cite{fischer03}; (4) \cite{udry06}; 
(5) \cite{wright09}; (6) \cite{butler97}; (7) \cite{butler99}; (8) \cite{butler06}; (9) \cite{tuomi13a} ; (10) \cite{queloz00}; (11) \cite{kurster00}; 
(12) \cite{mayor04}; (13) \cite{pepe11}; (14) \cite{hatzes00} ; (15) \cite{galland05}; (16) \cite{jones02}; (17) \cite{mayor09}; (18) \cite{lovis06}; 
(19) \cite{endl12}; (20) \cite{marcy02}; (21) \cite{mcarthur04}; (22) \cite{fischer08};
(23) \cite{gregory10}; (24) \cite{butler96}; (25) \cite{fischer02}; (26) \cite{marcy05}; (27) \cite{meschiari11}; 
(28) \cite{tinney11}; (29) \cite{vogt10}; (30) \cite{marcy96}; (31) \cite{brogi12}; (32) \cite{dumusque12}; (33) \cite{forveille11};
(34) \cite{bonfils05}; (35) \cite{udry07}; (36) \cite{noyes97}; (37) \cite{wittenmyer07}; (38) \cite{butler03}; 
(39) \cite{wright08}; (40) \cite{wright07}; (41) \cite{pepe07}; (42) \cite{butler01}; (43) \cite{mccarthy04};
(44) \cite{santos04}; (45) \cite{gozdziewski07}; (46) \cite{muterspaugh10}; (47) \cite{cochran97}; (48) \cite{mayor11}; 
(49) \cite{howard11}; (50) \cite{bailey09}; (51) \cite{marcy99}; (52) \cite{rivera10}; (53) \cite{delfosse98}; 
(54) \cite{marcy98}; (55) \cite{marcy01}; (56) \cite{rivera05}; (57) \cite{mayor95}; (58) \cite{fischer99}; (59) \cite{vogt05}.
\twocolumn

\end{document}